\documentclass[a4paper,11pt]{article}
\pdfoutput=1 % if your are submitting a pdflatex (i.e. if you have
\usepackage{jheppub} % for details on the use of the package, please
\usepackage[T1]{fontenc} % if needed

\usepackage{braket}
\usepackage[dvipsnames]{xcolor}
\usepackage{comment}
\usepackage{float}
\usepackage{caption}
\usepackage{subcaption}
\usepackage{bbold}  
\usepackage{hyperref}
\title{\boldmath JT gravity from non-Abelian T-duality}

\author[a]{Daniele Bielli}
\author[b]{Silvia Penati}
\author[b]{Anayeli Ramirez}

\affiliation[a]{High Energy Physics Research Unit, Faculty of Science,
Chulalongkorn University,
\\
Bangkok 10330, Thailand}
\affiliation[b]{Dipartimento di Fisica, Università degli studi di Milano–Bicocca,
\\
and INFN, Sezione di Milano–Bicocca,
\\
Piazza della Scienza 3, 20126 Milano, Italy}

\emailAdd{d.bielli4@gmail.com, silvia.penati@mib.infn.it, anayeli.ramirez@mib.infn.it}

\abstract{We study the geometries obtained by performing super non-Abelian T-duality of the Principal Chiral Model on OSp$(1|2)$. While the initial model represents an appropriate 3D supergravity background, interpretable as the superspace version of AdS$_{3}$, the T-dual model fails solving 
the 3D supergravity torsion constraints. We argue that this has to do with a factorisation pattern taking place under dualisation: the dual 3D geometry can be rewritten as the supersymmetric version of AdS$_{2}$, satisfying the supergravity constraints, fibered over what we interpret as the superspace equivalent of the standard bosonic line. We discuss an interesting connection between T-duals  of generic Principal Chiral Models and Poisson sigma models. We exploit it to show that in a suitable limit the dual action studied in this work gives rise to JT (super)gravity.}

\begin{document} 
\maketitle
\flushbottom

\section{Introduction}
\label{sec:intro}

Non-Abelian T-duality (NATD) was originally formulated for two-dimensional non-linear sigma models in \cite{delaOssa:1992vci, Giveon:1993ai,Alvarez:1993qi,Alvarez:1994np,Sfetsos:1994vz,Lozano:1995jx,Sfetsos:1996pm}, as a generalization of the well known Abelian T-duality \cite{Buscher:1987sk,Buscher:1987qj,Rocek:1991ps}. Since then, NATD
has developed as a very efficient technique for generating new supergravity solutions, starting from a given background geometry that exhibits a non-Abelian group of isometries \cite{Gasperini:1993nz,Gasperini:1994du} (useful reviews on this subject can be found in \cite{Itsios:2013wd,Hong:2018tlp,Thompson:2019ipl}). 
In string theory, this has allowed to find new supergravity backgrounds with AdS factors and typically reduced amounts of supersymmetry \cite{Sfetsos:2010uq,Lozano:2011kb,Lozano:2012au,Itsios:2012zv,Macpherson:2013zba,Lozano:2013oma,Zacarias:2014wta,Lozano:2014ata,Sfetsos:2014tza, Kelekci:2014ima,Kooner:2014cqa,Lozano:2015cra,Macpherson:2015tka,vanGorsel:2017goj,Ramirez:2021tkd}. In this context, most of the results come from dualising a compact group of isometries associated with some internal compact space. More recently, inspired by \cite{Alvarez:1993qi}, NATD has been generalised to the case of non-compact groups \cite{Ramirez:2021tkd}, where an AdS$_2$ solution was obtained as a by-product of the dualisation process. Nowadays, the studies involving AdS$_2$ geometries are motivated by developments in black hole physics and as AdS duals of strongly coupled fields theories \cite{Strominger:1998yg,Hartman:2008dq,Cadoni:1999ja,Balasubramanian:2009bg,Azeyanagi:2007bj,Castro:2014ima,Cvetic:2016eiv,Anninos:2008qb,Lozano:2020txg}, including the holographic description of one dimensional defects \cite{DHoker:2007mci,Chiodaroli:2009xh,Lozano:2020sae,Lozano:2021fkk,Lozano:2022vsv,Lozano:2022swp}.

The idea of extending the Abelian dualisation procedure to the case of backgrounds enjoying commuting superisometries was introduced in \cite{Berkovits:2008ic,Beisert:2008iq} and further developed in \cite{Hao:2009hw,Adam:2009kt,Adam:2010hh,Dekel:2011qw,Bakhmatov:2010fp,Abbott:2015mla,Colgain:2016gdj} - see also \cite{Ricci:2007eq} for previous related work and \cite{OColgain:2012si} for a review. In recent years, it has led to various non-Abelian generalisations, formulated in terms of BRST techniques \cite{Grassi:2011zf}, super Poisson-Lie symmetry \cite{Eghbali:2009cp,Eghbali:2011su,Eghbali:2012sr,Eghbali:2013bda,Eghbali:2014coa,Eghbali:2020twu,Eghbali:2023sak,Eghbali:2024vxi}, Double Field Theory formalism \cite{Astrakhantsev:2021rhj,Astrakhantsev:2022mfs,Astrakhantsev:2023urj,Butter:2023nxm}, as well as manifestly supersymmetric generalizations of the NATD technique of de la Ossa and Quevedo \cite{Borsato:2016pas,Borsato:2017qsx,Borsato:2018idb,Borsato:2021vfy,Bielli:2021hzg,Bielli:2022gmm,Bielli:2023bnh}, which applies to supersymmetric non-linear sigma models formulated directly in superspace. The latter relies on the existence of a supergroup of isometries in the original supermanifold and leads to dual supergeometries where supersymmetry is generally reduced. This property is a direct result of the non-Abelian dualisation procedure, which has the feature of preserving only those isometries which commute with the gauged ones \cite{Plauschinn:2013wta,Bugden:2018pzv}.

In \cite{Bielli:2021hzg}, the application of super NATD (SNATD) has been worked out in detail for the Principal Chiral Model (PCM) on OSp$(1|2)$. This model describes a $\mathcal{N}=1$ supersymmetric generalisation of AdS$_3$ (sAdS$_3$), and allows for a consistent description as a supermanifold satisfying the torsion constraints of 3D superspace \cite{Gates:1983nr,
Buchbinder:1998qv,Buchbinder:2017qls}. Dualising the left sector of the  OSp$(1|2)_L \times$ OSp$(1|2)_{R}$ isometry group, one lands on a dual geometry whose isometries are reduced to the non-dualised right sector.
However, as noted in \cite{Bielli:2021hzg}, the supervielbeins of the dual model no longer satisfy supergravity constraints, though the background still exhibits manifest supersymmetry. The situation does not improve upon dualising only the maximal bosonic subgroup of OSp$(1|2)_{L}$ - analogous results would be obtained by performing dualisation in the right sector OSp$(1|2)_{R}$.
This problem was left open, with the suggestion that a possible consistent embedding of the dual 3D geometry inside a 10D string solution might cure this pathology. 

In this paper we revisit such a problem, inspired by what happens in the purely bosonic case, that is in the dualisation of the 
AdS$_3$ background with no fluxes. This is notoriously described by a PCM on SL$(2, {\mathbb R})$. Performing NATD along the left sector SL$(2, {\mathbb R})_{L}$ of the isometry group \cite{Alvarez:1993qi}, the resulting dual model turns out to have a non-vanishing $B$ field, but the corresponding metric is no longer a solution of the 3D Einstein equations coupled to $B$\footnote{This does not contradict general expectations, as it is well known that NATD is not a mapping between Einstein solutions, i.e. it does not generally map solutions to solutions within the pure Einstein sector. Indeed, if we think of AdS$_3$ geometry as a consistent truncation of a 10D supergravity background, extra fields arise and this issue disappears \cite{Ramirez:2021tkd}. The 3D background should now solve 3D gauged supergravity equations with additional fields.}. Rather, in a suitable set of coordinates it exhibits a warped factorisation in terms of an AdS$_2$ and a half-line (see equation \eqref{warp}) \cite{Ramirez:2021tkd}, where the 2D part correctly satisfies pure Einstein equations (on the 2D slice $H = {\textrm d}B$ vanishes). 
Somehow, under NATD a natural dimensional reduction from AdS$_3$ to AdS$_2$ occurs. 

Here, we show that something similar occurs in the supersymmetric case. The dual supergeometry, no longer a solution of 3D supergravity constraints, decomposes into a super-AdS$_2$ factor, satisfying the constraints in two dimensions, and the supersymmetric version of a 1d line element - namely a (1|2)-dimensional supermanifold that we will call \textit{superline} (see equation \eqref{eq:sol2}). Therefore, in analogy with the bosonic model, SNATD induces a natural dimensional reduction from three to two dimensions. However, there is an important difference between the two cases. While in the bosonic dual model there is a complete factorization between the AdS$_2$ and the half-line, and the warp factor in front of the AdS$_2$ metric depends only on the half-line coordinate, in superspace the sAdS$_2$ metric, the superline element and the warp factor share the same spinorial coordinates. Factorization in the spinorial sector is prevented by three-dimensional supersymmetry, which in particular requires that the variation of the superline element compensates for the non-trivial variation of the warp factor in \eqref{eq:sol2}, while sAdS$_2$ is invariant.  

To further support the  interpretation that in the studied model (S)NATD acts as a dimensional-reduction-inducing procedure, we show that in a suitable limit the action of the three-dimensional dual model  reduces to a Jackiw-Teitelboim-like (super)gravity action in two dimensions, both in the bosonic and the supersymmetric cases. 

It is known that in two dimensions a non-trivial formulation of gravity requires the  introduction of a dilaton field. The action was first proposed by Jackiw \cite{Jackiw:1984je} and Teitelboim \cite{Teitelboim:1983ux}, and successively generalized to supergravity \cite{Teitelboim:1983uy,Chamseddine:1991fg,Izquierdo:1998hg,Ikeda:1993fh,Ikeda:1993dr}. In particular, a dilaton-gravity action with an algebraic dilaton subject to a constant potential describes AdS$_2$ geometry, while its ${\cal N}=1$ supersymmetric generalization gives sAdS$_2$. The AdS$_2$ action can be equivalently formulated as a Poisson sigma model in three dimensions, with the dilaton field playing the role of the third direction  \cite{Schaller:1994es,Schaller:1994pm}. The same is true for the sAdS$_2$ action, which can be identified with a three-dimensional graded Poisson sigma model  \cite{Strobl:1999zz, Ertl:2000si,Bergamin:2002ju}. 

Given the warped factorization of our dual models into (s)AdS$_2$ times a (super)line, it is then tempting to conjecture that the (S)NATD action may reduce to the JT (super)gravity action with the third direction somehow identified with the dilaton. To this end, we show that introducing a suitable coupling constant in the dualisation procedure, in the small coupling limit the three-dimensional dual action can be rewritten as a Poisson sigma model. Therefore, identifying one of the target space coordinates with the dilaton, the dual action reduces to (s)AdS$_2$ JT (super)gravity. In conclusion, NATD and SNATD applied to three-dimensional sigma models describing AdS$_3$ and sAdS$_3$ geometries lead naturally to JT (super)gravity actions, consistently with the observed foliation of the geometries in terms of AdS$_2$ and sAdS$_2$ slices, respectively. 

The rest of the paper is organized as follows. In section \ref{sec:review} we briefly review the (S)NATD procedure applied to the PCMs on SL$(2, {\mathbb R})$  and OSp$(1|2)$, recalling in particular the warp factorization of dual metric in terms of an AdS$_2$ slice plus a half-line, which occurs in the bosonic  case. The analogue factorization arising in the supersymmetric case is investigated in section \ref{sec:sAdS2}, where a detailed description of the non-trivial changes of coordinates necessary to dig out the $\mathcal{N}=1$ sAdS$_2$ factor is provided. Section \ref{sec:JT} is devoted to showing how the dual actions, both the bosonic and the supersymmetric ones, can be reduced to two-dimensional JT (super)gravity actions, thus supporting the idea that the factorization which arises under (S)NATD may have a deeper interpretation. Finally, in section \ref{sec:discussion} we close with a discussion on possible applications and future developments. 
The paper is complemented by two appendices that collect our conventions and details of the calculations.

\section{(Super)NATD of Principal Chiral Models}\label{sec:review}
We  begin with a short review of (S)NATD applied to  Principal Chiral Models defined on (super)group manifolds G. In terms of the Maurer-Cartan form $j=\texttt{g}^{-1}\mathrm{d}\texttt{g}$ with $\texttt{g}: \Sigma \rightarrow$ G being a (super)group element, the PCM action reads 
\begin{equation}\label{PCM-action}
S_{PCM} = -\tfrac{1}{2} \int_{\Sigma} \langle j , \star j \rangle \, ,
\end{equation}
where $\langle \cdot, \cdot \rangle$ is a non-degenerate, Ad-invariant and (graded-)symmetric bilinear form on the Lie (super)algebra $\mathfrak{g}$, while $\star$ is the Hodge operator with respect to the worldsheet metric. 

Parametrizing the (super)group element $\texttt{g}$ in terms of $Z^{A} \equiv \{ z^{a}, \chi^{\alpha} \}$ coordinates, with $a \in\{ 0,...,d_{b}-1  \}$ and $\alpha \in \{ 1,..., d_{f} \}$, the above action can be recast in the following form\footnote{Along the paper we will be using the symbol $Z^A$ to indicate the coordinates of the original model and $X^A$ for the ones of the dual model. Similarly, $\hat G, \hat B$ will refer to the geometrical tensors of the original model, leaving $G, B$ for the dual ones. The rest of our conventions are collected in appendix \ref{appendix:osp-algebra}.}
\begin{equation}
\label{PCM-action2}
  S_{PCM}   = \int_{\Sigma} \mathrm{d}\tau\mathrm{d}\sigma \, \, 
 \sqrt{-h} \, h^{ij} \,\partial_{j} Z^{B}\partial_{i}Z^{A} \, \hat G_{AB} \, ,
\end{equation}
where $\tau,\sigma$ denote coordinates on the Lorentzian worldsheet $\Sigma$, equipped with metric $h_{ij}$, and the overall $-1/2$ factor in \eqref{PCM-action} has been absorbed in $\hat G$.

The model encodes a $(d_{b}|d_{f})$ background (super-)geometry described by a graded-symmetric metric $\hat G_{AB}=(-1)^{AB}\hat{G}_{BA}$, with $d_{b}$ and $d_{f}$ denoting its bosonic and fermionic dimensions, respectively.  It exhibits G$_{L}\times$G$_{R}$ isometry as a result of the invariance of \eqref{PCM-action} under global transformations of the form $\texttt{g} \rightarrow g_{L}^{-1} \, \texttt{g} \, g_{R}$, with $g_{L},g_{R} \in$ G.

T-dualisation with respect to a given subgroup H of the isometry group, for concreteness taken to be H $\subseteq$ G$_{L}$, requires first the gauging of H \cite{delaOssa:1992vci,Rocek:1991ps}. This  modifies the Maurer-Cartan form as $j_{\omega} \equiv \texttt{g}^{-1}(\mathrm{d}+\omega)\texttt{g}$, with $\omega \in \mathfrak{h}=$Lie(H). The gauge fields transform as $\omega \rightarrow h^{-1}\omega h + h^{-1}\mathrm{d}h$, leaving $j_{\omega}$ and the gauged action invariant under the local transformations $\texttt{g}\rightarrow h^{-1}\texttt{g}$ with $h\in$ H. To recover the initial model, the introduction of gauge fields must be accompanied by the enforcement of their flatness $F_{\omega}\equiv \mathrm{d}\omega +\tfrac{1}{2}[\omega,\omega]=0$. This is achieved by using Lagrange multipliers $\Lambda \in \mathfrak{h}$, which transform as $\Lambda \rightarrow h^{-1}\Lambda h$ for invariance of the full \textit{master action}
\begin{equation}\label{master-action}
S_{\omega} = -\tfrac{1}{2} \int_{\Sigma} \langle j_{\omega}, \star j_{\omega} \rangle - \int_{\Sigma} \langle \Lambda, F_{\omega} \rangle \, .
\end{equation}
Integrating out the Lagrange multipliers enforces the flatness of $\omega$ and, at least under the assumption of a topologically trivial worldsheet, allows to set $\omega = 0$, thus recovering the initial model \eqref{PCM-action}. On the other hand, integrating out the gauge fields we obtain the T-dual sigma model in which the Lagrange multipliers play the role of background coordinates. Its action can be written in the form \cite{Bielli:2021hzg,Bielli:2023bnh}
\begin{equation}\label{T-dual-sigma model-paper-notation}
\tilde S=-\frac{1}{2} \int_{\Sigma} \,\, \langle \textrm{d} X, \frac{1}{1-\textrm{ad}_X} (1+ \star )\textrm{d} X \rangle, \qquad \text{with} \qquad 
X \equiv \texttt{g}^{-1}\Lambda \texttt{g} = X^A T_A  \, , 
\end{equation}
where $T_{A}$ are the generators of $\mathfrak{g}$ and $\textrm{ad}_{X}(Y) \equiv [X,Y]$, with $\textrm{ad}_{X}^0\equiv 1$. As described in appendix \ref{appendix:osp-algebra}, the action can be conveniently rewritten as
\begin{equation}\label{general-T-dual-sigma-model}
\tilde S=\int_{\Sigma} \mathrm{d}\tau\mathrm{d}\sigma  
\left(  \sqrt{-h} \, h^{ij} \,\partial_{j} X^{B}\partial_{i} X^{A} \, G_{AB} + \varepsilon^{ij} \, \partial_{j} X^{B} \partial_{i} X^{A} \, B_{AB} \right) \, ,
\end{equation}
where, up to an overall $-1/2$ factor, the dual metric $G$ and the $B$ field are identified with the symmetric and antisymmetric parts of $\frac{1}{1-\mathrm{ad}_X}=\sum_{k=0}^\infty \mathrm{ad}_X^k$. In components, these are graded symmetric $G_{AB}=(-1)^{AB}G_{BA}$ and graded antisymmetric $B_{AB}=-(-1)^{AB}B_{BA}$.

The dual coordinates $X^A \equiv \{ x^a , \theta^{\alpha} \}$ are a mixture of Lagrange multipliers $\Lambda$ and  coordinates $Z$ of the original model, depending on the choice of the gauged subgroup H. In this work, we will consider the two cases H $=$ G$_{L}$ and H $=$ H$_{bos}$, the latter representing the maximal bosonic subgroup of the supergroup G$_{L}$. In the first case, gauge invariance can be exploited to get rid of the original coordinates completely, that is setting $\texttt{g}=\mathbb{1}$. Consequently, the dual coordinates $X^{A}$ coincide with the Lagrange multipliers $\Lambda^A$. In the second case instead, only the initial bosonic coordinates can be removed and replaced by the Lagrange multipliers, whereas the fermionic coordinates are still those of the initial model \eqref{PCM-action2}. In both cases only the right sector G$_{R}$ of the isometry group, which commutes with the gauged left sector, survives in the dual model.

For the sake of completeness it should be also mentioned that while the above procedure represents a classically valid way of performing T-duality, quantum corrections might be taken into account by including a shift in the dilaton \cite{delaOssa:1992vci,DeJaegher:1998pp,Berkovits:2008ic,Hollowood:2014qma, Hoare:2015gda, Borsato:2016ose}.
This is proportional to the inverse of the  determinant of the (super)matrix 
\begin{equation}
\label{eq:supermatrixN}
N_{AB}=\delta_{AB}-\Lambda^C f_{CAB} \, ,
\end{equation}
where $f_{ABC}$ are the lowered-index structure constants of the Lie superalgebra.

In the rest of the paper we focus on the PCM on the supergroup manifold OSp$(1|2)$, which describes a $(3|2)$-dimensional background with OSp$(1|2)_{L} \times$OSp$(1|2)_{R}$ isometry group. It can be regarded as the supersymmetric generalization of the AdS$_{3}$ PCM based on the SL$(2,\mathbb{R})_{L} \times$ SL$(2,\mathbb{R})_{R}$ group. 
The sAdS$_3$ model can be shown to satisfy the torsion constraints required for a physical interpretation of superspaces in terms of supergravity theories \cite{Kuzenko:2011rd, Kuzenko:2012bc}. 

\subsection{NATD of the PCM on SL$(2,\mathbb{R})$}
Non-Abelian T-duality of the PCM on SL$(2,\mathbb{R})$, describing the AdS$_{3}$ background, was performed in \cite{Alvarez:1993qi} by dualising the left sector SL$(2,\mathbb{R})_L$ of the isometry group. The dual model is featured by a metric and a non-trivial $B$ field, that in terms of dual coordinates $x^a, a \in \{ 0,1,2 \}$, read
\begin{equation}
\label{eq:dualbosonic}
    \mathrm{d}s^2 = - \frac{1}{4(r^2 -1)} (\eta_{ab} - x_a x_b ) \, \mathrm{d}x^a \, \mathrm{d}x^b \, , \qquad \quad B =  \frac{1}{4(r^2 -1)} \, \varepsilon_{abc} x^c \, \mathrm{d}x^a \wedge \mathrm{d}x^b \, ,
\end{equation}
where we have defined $r^2 \equiv \eta_{ab} \, x^a x^b$. Without loss of generality, we will always restrict to the slice $r^2>0$. For $r^2 <0$ the resulting geometry is of the same form but exhibits a singularity at $r^2=-1$. For details we refer to \cite{Alvarez:1993qi,Ramirez:2021tkd,Lozano:2022fqk}.

In addition, if we perform the duality transformation at the quantum level, a dilaton field arises, which is given by $e^{-2\Phi}=r^2-1$.

The expressions in \eqref{eq:dualbosonic} seem to elude a geometrical interpretation, as they exhibit a physical singularity at $r^2 = 1$ and the metric does not satisfy 3D Einstein equations coupled to $B$. However, as discovered later  \cite{Lozano:2021rmk,Ramirez:2021tkd}, upon a suitable change of dual coordinates $\{\ \! \! x^0, x^1, x^2 \} \to \{ t, 
u, r \equiv \sqrt{r^2}\}$, defined as
\begin{equation}
\label{eq:paramet}
\{x^0,x^1,x^2\}=\left\{ - \frac{ru}{2}  \left( 1- \frac{1+t^2}{u^2} \right),\; \frac{ru}{2}  \left( 1+ \frac{1-t^2}{u^2} \right),\; \frac{rt}{u} \right\} \, ,
\end{equation}
the dual metric and the $B$ field can be recast into the following form
\begin{equation}\label{eq:factorization}
\mathrm{d}s^2  = \Omega(r) \, \mathrm{d}s^2_{\rm AdS_2} + \mathrm{d}r^2  \qquad \qquad \qquad 
B =r \, \Omega(r) \, {\rm vol}_{\text{AdS}_2} 
\end{equation}
with the AdS$_{2}$ metric, volume form and warping factor reading
\begin{equation}\label{warp}
\mathrm{d}s^2_{\rm AdS_2} = \frac{1}{u^2} (-\mathrm{d}t^2 +\mathrm{d}u^2) \qquad \qquad \text{vol}_{\text{AdS}_2}=\frac{1}{u^2} \mathrm{d}t\wedge \mathrm{d}u
\qquad \qquad 
\Omega(r) = \frac{r^2}{r^2-1} \,\, .
\end{equation}
 Under NATD the original AdS$_3$ background gets replaced by a warped AdS$_2 \times {\mathbb R}^+$ space, where the AdS$_2$ factor contains the SL$(2, \mathbb{R})$ residual symmetry. Though AdS$_2$ is a solution of 2D pure gravity equations, the 3D equations of motion are spoiled by the appearance of extra fields and a warp factor $\Omega(r)$, which is no longer constant when we uplift AdS$_2$ to three dimensions by turning on the $r$ coordinate. We refer to \cite{Itsios:2012zv} for a similar pattern, which arises when  dualising compact spaces.

 \vskip 10pt

For the sAdS$_3$ model, SNATD was developed in \cite{Bielli:2021hzg, Bielli:2022gmm} dualising either the full left sector OSp$(1|2)_{L}$ of the isometry group, or on its maximal bosonic subgroup SL$(2,\mathbb{R})_{L}$. It turns out that the dual model has no longer the interpretation of a supergravity background in 3D superspace, as one lands on a dual sigma model which does not satisfy the supergravity torsion constraints. In addition, the dual metric exhibits a physical singularity which resembles the one of the bosonic case.  
It is thus tempting to investigate whether the SNATD of the original sAdS$_3$ geometry contains a meaningful two dimensional subspace, as it occurs in \eqref{warp} for the bosonic case.

In the next subsection we briefly revisit the results of \cite{Bielli:2021hzg, Bielli:2022gmm}
for the dual metric and $B$ field, both for the  OSp$(1|2)_L$ and SL$(2,\mathbb{R})_{L}$ dualisations, re-expressing them in a set of coordinates that is more suitable for the subsequent discussion on the geometric structure of the dual supermanifold. 
We will adopt vector notation, as done in \cite{Bielli:2022gmm}. A summary of conventions can be found in appendix \ref{appendix:osp-algebra}.

\subsection{Super NATD of the PCM on OSp(1|2)}
\label{sec:NATD of OSp(1|2) supergroup}

The super non-Abelian T-dual model to the PCM on OSp(1|2) describe a $(3|2)$ dimensional supermanifold, coordinated by $X^A = \{x^a, \theta^\alpha \}$, with $a \in \{ 0,1,2 \}$ and $\alpha \in \{ 1,2 \}$. 
 The corresponding metric and $B$ field in \eqref{general-T-dual-sigma-model} are supermatrices of the form 
\begin{equation}
    G_{AB} = \begin{pmatrix}
        G_{ab} & G_{a \beta} \\
        G_{\alpha  b}  & G_{\alpha \beta} 
    \end{pmatrix} \, , \qquad \qquad 
    B_{AB} = \begin{pmatrix}
        B_{ab} & B_{a \beta} \\
        B_{\alpha  b}  & B_{\alpha \beta} 
    \end{pmatrix} \, ,
\end{equation}
where $G_{ab}$ is the usual symmetric tensor, spinorial mixed components are also symmetric, whereas double-spinor index components satisfy $G_{\alpha \beta} = - G_{\beta\alpha}$. Similarly, $B_{ab}$ are the usual antisymmetric bosonic components, $B_{a\beta}$ are also antisymmetric while the double-spinor index components are symmetric. 

Performing SNATD of the full left sector OSp$(1|2)_{L}\subseteq$OSp$(1|2)_{L}\times$OSp$(1|2)_{R}$ of the isometry supergroup, one obtains the following components \cite{Bielli:2021hzg, Bielli:2023bnh} 
\begin{equation} \label{eq:GBcomponents}
\begin{aligned}
G_{ab}& =-\tfrac{1}{2}L_{1}(r) \, \bigl[ \eta_{ab}\bigl(1-2i \, f(r) \, \theta^{1}\theta^{2}\bigr)-x_{a}x_{b}\bigl( 1-2i \, g(r) \, \theta^{1}\theta^2\bigr) \bigr],
\\
G_{\alpha \beta}&=\tfrac{1}{2}L_{2}(r) \, \epsilon_{\alpha \beta}\bigl(1-2i\, h(r)\, \theta^{1}\theta^{2}\bigr),
\\
G_{a \alpha}&=-\tfrac{1}{2}L_{3}(r) \, \bigl[ -(1-r^2)x_{a}\epsilon_{\alpha\beta}+3\varepsilon_{abc}(\gamma^{b})_{\alpha \beta} \, x^{c} \bigr]\theta^{\beta},
\\
\\
B_{ab}&=-\tfrac{1}{2}L_{1}(r) \, \varepsilon_{abc}\, x^{c}\bigl( 1-2i\, m(r)\, \theta^{1}\theta^{2}\bigr), 
\\
B_{\alpha \beta}&=\tfrac{1}{4}L_{2}(r) \, x_{a}(\gamma^{a})_{\alpha \beta}\bigl(1-2i\, n(r)\, \theta^{1}\theta^{2}\bigr),
\\
B_{a\alpha}&=-\tfrac{1}{2}L_{3}(r) \, \bigl[ (2+r^2)(\gamma_{a})_{\alpha\beta}-3x_{a}x_{b}(\gamma^{b})_{\alpha\beta} \bigr]\theta^{\beta} \, , 
\end{aligned}
\end{equation}
with the $r$-dependent functions explicitly given by ($r^2>0$)
\begin{align}
\label{eq:Ls}
& \! L_{1}(r)=\frac{1}{2(r^2-1)}, 
&& \qquad L_{2}(r)=\frac{4i}{(r^2-4)},
&&& \; L_{3}(r)=-\frac{i}{(r^2-1)(r^2-4)},
\notag \\
& \; f(r) =\frac{2(2r^2+1)}{(r^2-1)(r^2-4)},
&& \qquad  \; \; g(r)= -\frac{2}{(r^2-1)},
&&&h(r) = \frac{r^4-7r^2-12}{4(r^2-1)(r^2-4)},
\notag \\
&m(r)= \frac{r^2+5}{(r^2-1)(r^2-4)},
&& \qquad  \; \; n(r) = \frac{r^2-10}{(r^2-1)(r^2-4)} \, \, .
\end{align}

\vskip 5pt
For our purposes, it is convenient to parametrize the bosonic sector in terms of the new set of coordinates $\{t,u,r \}$ defined in \eqref{eq:paramet}, and perform the following rescaling of the fermionic coordinates
\begin{equation}\label{susy-dualisation-fermionic-rescaling}
 \theta^\alpha \rightarrow \sqrt{\frac{r^2 - 4}{6}}  \, \theta^\alpha \, .   
\end{equation}
Moreover, the fermionic sector is better organized using the following one-form 
\begin{align}\label{eq:A}
   \mathcal{A}&=  \frac{1}{u} \left[ (t^2-u^2)\, \mathrm{d}\theta^1 \theta^1+ \, \mathrm{d}\theta^2 \theta^2+t \, (\mathrm{d}\theta^1\theta^2+\mathrm{d}\theta^2 \theta^1) \right],
\end{align}
its derivatives 
\begin{equation}
\begin{aligned}
\label{eq:Aderivatives}
\partial_t \mathcal{A}&=\frac{1}{u} (2t \, \mathrm{d}\theta^1 \theta^1+\mathrm{d}\theta^1 \theta^2+\mathrm{d}\theta^2 \theta^1), 
\\
\partial_u \mathcal{A}&= - \frac{1}{u^2} \left[ (u^2+t^2) \, \mathrm{d}\theta^1 \theta^1+\mathrm{d}\theta^2 \theta^2+t \, (\mathrm{d}\theta^1 \theta^2+\mathrm{d}\theta^2 \theta^1) \right]\, ,
\end{aligned}
\end{equation}
and its exterior derivative, $\mathrm{d}\mathcal{A} \equiv \mathrm{d}_B\mathcal{A}+\mathrm{d}_F\mathcal{A}$, with
\begin{equation}
\begin{aligned}
\label{eq:derA}
&\mathrm{d}_B\mathcal{A}=\mathrm{d}\left[\frac{1}{u} (t^2-u^2)\right]\wedge \mathrm{d}\theta^1 \theta^1 - \frac{\mathrm{d}u}{u^2}\wedge \mathrm{d}\theta^2 \theta^2 +\mathrm{d}\left[\frac{t}{u}\right]\wedge(\mathrm{d}\theta^2 \theta^1+\mathrm{d}\theta^1\theta^2) \, ,
\\
&\mathrm{d}_F\mathcal{A}= - \frac{1}{u} \left[ (t^2-u^2)\, \mathrm{d}\theta^1\wedge \mathrm{d}\theta^1+\, \mathrm{d}\theta^2\wedge \mathrm{d}\theta^2+2\, t\, \mathrm{d}\theta^1\wedge \mathrm{d} \theta^2 \right] \, .
\end{aligned}
\end{equation}
In terms of the new coordinates $\{t,u,r \}$ and the $\cal A$ form above, the T-dual line element and the $B$ field read 
\begin{align}
\mathrm{d}s^2 
&= \, \frac{r^2}{4(r^2-1)} \left(1-\frac{2i(2r^2+1)}{3(r^2-1)} \, \theta^1\theta^2\right)\mathrm{d}s^2_{\text{AdS}_2}
\label{eq:PCDMPoincaremetric} \\
& +  \frac{1}{4}\left(1-\frac{2i}{3}\frac{r^4 - 4r^2 - 16}{(r^2-4)^2}\theta^1\theta^2\right)\mathrm{d}r^2-\frac{ir(r^2-8)}{6(r^2-4)} \mathrm{d}r \, \mathrm{d}(\theta^1\theta^2) 
\notag \\
&+\frac{2i}{3}\left(1-i\frac{r^4-7r^2-12}{12(r^2-1)} \, \theta^1\theta^2\right)\mathrm{d}\theta^1 \mathrm{d}\theta^2 -\frac{ir^2}{2(r^2-1)}\left(\mathrm{d}u \, \partial_t\mathcal{A} +\mathrm{d}t \, \partial_u\mathcal{A}\right) \, ,
\notag\\
\notag \\
B&=\frac{r^3}{2(r^2-1)} \left(1- \,  \frac{i(r^2+5)}{3(r^2-1)} \, \theta^1\theta^2\right)
{\rm vol}_{\text{AdS}_2} 
\label{eq:PCDMPoincareB2} \\
& -\frac{4i}{3(r^2-4)} \mathrm{d}r\wedge \mathcal{A} -\frac{ir(r^2+2)}{6(r^2-1)} \;\mathrm{d}_B\mathcal{A} -\frac{ir}{6}\left(1-\frac{i(r^2-10)}{3(r^2-1)} \, \theta^1\theta^2\right)\mathrm{d}_F\mathcal{A} \, , 
\notag
\end{align}
with $\mathrm{d}s^2_{\rm AdS_2}$ and vol$_{\text{AdS}_2}$ defined in \eqref{warp}. A dilaton shift is also produced, which reads
\begin{equation}\label{eq:Dilaton-poincare-snatd2}
    e^{-2\Phi}=  \frac{1}{2(r^2-4)} \left(r^2-1+4i \, \theta^1\theta^2 \right) \, .
\end{equation}
We stress that, while the T-dual geometry of the bosonic PCM has 
a physical singularity at $r^2=1$, in the OSp(1|2) case the dual geometry exhibits an additional  singularity at $r^2=4$ ($r=2$ in the region of interest). This is entirely encoded in the fermionic sector of the supermanifold, as it appears in (\ref{eq:GBcomponents}, \ref{eq:PCDMPoincaremetric}, \ref{eq:PCDMPoincareB2}) multiplied by the $\theta$-coordinates, and indeed results from the dualisation of superisometries. In \eqref{eq:Dilaton-poincare-snatd2} it originates from the determinant of the fermion-fermion block contained in the $N_{AB}$ supermatrix in \eqref{eq:supermatrixN}. 
It then follows that physically meaningful quantities, like the purely bosonic Riemann tensor and Ricci scalar, are not affected by this singularity, which can be then interpreted as a coordinate singularity.

If we turn off the fermionic coordinates, expressions (\ref{eq:PCDMPoincaremetric}, \ref{eq:PCDMPoincareB2}) correctly reduce to the ones for the metric and the $B$ field obtained from the T-dualisation of the PCM on SL$(2,\mathbb{R})$, in the form \eqref{eq:factorization}.  
In particular, we recover the bosonic AdS$_2 \times \mathbb{R}^+$ geometry.
The reduction of the dilaton field requires, in addition, to remove the contribution to the inverse of the superdeterminant of $N_{AB}$ in \eqref{eq:supermatrixN} coming from the fermion-fermion diagonal block $N_{\alpha\beta}$. This correctly removes the $1/(r^2-4)$ factor in \eqref{eq:Dilaton-poincare-snatd2}.

\vskip 15pt

We close this section by briefly recalling the results obtained by dualising the maximal bosonic subgroup SL$(2,\mathbb{R})_{L} \subseteq 
{\rm OSp}(1|2)_{L} \times {\rm OSp}(1|2)_{R}$ \cite{Bielli:2021hzg, Bielli:2023bnh}. 

The dual fields resulting from the procedure read
\begin{equation}
\begin{alignedat}{3}
&G_{ab}&&=-\tfrac{1}{2}L_{1}(r) \, \bigl( \eta_{ab}-x_{a}x_{b} \bigr), 
&& \qquad B_{ab}=-\tfrac{1}{2}L_{1} \, (r)\varepsilon_{abc} \, x^{c}, 
 \\
 &G_{\alpha \beta}&&=-\tfrac{i}{2} \epsilon_{\alpha \beta}\bigl(1-2iL_{2}(r) \, \theta^1\theta^2\bigr) ,
&& \qquad B_{\alpha\beta} \; =  -\tfrac{1}{4}L_{1}(r) \, x_{a}(\gamma^{a})_{\alpha \beta} \, \theta^1\theta^2 ,
\\
&G_{a\alpha}&&=-\tfrac{1}{2}L_{3}(r) \, \varepsilon_{abc}(\gamma^{b})_{\alpha\beta} \, x^{c} \,  \theta^\beta, 
&& \qquad B_{a\alpha}=-\tfrac{1}{2}L_{3} \, (r)\bigl(\eta_{ab}-x_{a}x_{b}\bigr)(\gamma^{b})_{\alpha\beta}\theta^\beta  \, ,
\end{alignedat}
\end{equation}
where the $L$ functions are given by
\begin{equation}
\label{eq:bosonic-dualisation-Ls}
L_{1}(r)=\frac{1}{2(r^2-1)}, 
\qquad \qquad
L_{2}(r)=-\frac{(r^2-2)}{8(r^2-1)},
\qquad \qquad
L_{3}(r)=\frac{i}{4(r^2-1)} \,\, ,
\end{equation}
and the fermionic coordinates $\theta^\alpha$ are still those of the original PCM. Using again the parametrisation \eqref{eq:paramet} we obtain the following T-dual metric, $B$-field and dilaton
\begin{equation}
\begin{aligned}
\label{eq:metric-poincare-sl2-natd} 
\mathrm{d}s^2& =  \frac{r^2}{4(r^2-1)} \, \mathrm{d}s^2_{\text{AdS}_2}
\\
& +\frac{1}{4} \, \mathrm{d}r^2 -i\left(1+\frac{i(r^2-2)}{4(r^2-1)} \, \theta^1\theta^2\right)\mathrm{d}\theta^1 \mathrm{d}\theta^2 +\frac{ir^2}{4(r^2-1)} \, \left(\mathrm{d}u \, \partial_t\mathcal{A}+ \, \mathrm{d}t \, \partial_u\mathcal{A}\right),
\\
\\
B& =\frac{r^3}{2(r^2-1)} \, 
{\rm vol}_{\text{AdS}_2}
\\
& +\frac{r}{8(r^2-1)} \, \theta^1\theta^2 \mathrm{d}_F\mathcal{A}+\frac{i}{4(r^2-1)}\left(r \, \mathrm{d}_B\mathcal{A}-(r^2-1) \, \mathrm{d}r\wedge \mathcal{A}\right),
\\
\\
 e^{-2\Phi}& = \, \frac12 (r^2 -1),
\end{aligned}
\end{equation}
with $\mathrm{d}s^2_{\rm AdS_2}$ and vol$_{\text{AdS}_2}$ defined in \eqref{warp} and $\mathcal{A},\mathrm{d}_{B}\mathcal{A},\mathrm{d}_{F}\mathcal{A}$ defined in (\ref{eq:A}, \ref{eq:derA}).

Once again, the bosonic geometry is AdS$_2\times\mathbb{R}^+$ and setting to zero the spinorial coordinates we are back to \eqref{eq:factorization}.
As anticipated above, in this case the coordinate singularity at $r^2=4$ does not arise, since dualisation only involves bosonic isometries.

\section{superAdS$_2 \times {\mathcal M^{(1|2)}}$ geometry from OSp(1|2) dualisation}\label{sec:sAdS2}

In the previous section we have identified an AdS$_2$ factor inside the bosonic part of the dual supermanifold, which resembles the dual geometry \eqref{warp} of the bosonic case. As we are going to discuss here, in the supersymmetric case we can do more. We can identify a super-AdS$_2$  submanifold, thus obtaining a supersymmetric version of \eqref{eq:factorization}. Since this requires an explicit identification of the sAdS$_2$ geometry, which to the best of our knowledge is not immediately available in the literature, and a non-trivial change of coordinates in the spinorial sector, able to make this clearly visible, we shall describe the procedure in detail. To begin with, we review the sAdS$_2$ geometry. 

\subsection{Identifying sAdS$_2$}
\label{subsec:Identifying-AdS2}
 
Following \cite{Verlinde:2004gt,Grassi:2005kc,McGuigan:2004sq}, we consider a $\kappa$-symmetric Green-Schwarz action for a  semi-symmetric coset OSp(1|2)/SO(1,1), which in our conventions reads
(see eq. (10) in \cite{Verlinde:2004gt})
\begin{equation}
S=-\int_{\Sigma}d^2 z \,\, \frac{(\bar{\partial}Z+i\Theta \bar{\partial}\Theta)(\partial \bar{Z}+i\bar{\Theta}\partial \bar{\Theta})}{(Z-\bar{Z}-i\Theta\bar{\Theta})^2} \, .
\end{equation}
Here $Z,\bar{Z},\Theta,\bar{\Theta}$ are real light-cone super-coordinates for the $(2|2)$-dimensional background, whereas $z,\bar{z}$ are real light-cone coordinates on the worldsheet. 

In order to recast the above action in the standard Green-Schwarz form, we first go back to $(\tau, \sigma)$ worldsheet coordinates by setting,
$z=\tfrac{1}{2}(\tau+\sigma),\; \bar{z}=\tfrac{1}{2}(\tau-\sigma)$, and redefine
\begin{equation}
Z=\frac{1}{2}(t+u ), \qquad \bar{Z}=\frac{1}{2}(t - u), \qquad \Theta=\frac{1}{\sqrt{2}}(\tilde{\theta}^1+\tilde{\theta}^2), \qquad \bar{\Theta}=\frac{1}{\sqrt{2}}(\tilde{\theta}^1-\tilde{\theta}^2) \,\, .
\end{equation}
As a result, the sAdS$_2$ action takes the form \eqref{general-T-dual-sigma-model}, with the metric and $B$ field components explicitly given by
\begin{align}
\label{eq:dsVerlinde}
\mathrm{d}s^2_{\text{sAdS}_2}=&\frac{1}{8}(1 -\frac{2i}{u} \tilde{\theta}^1 \tilde{\theta}^2)\mathrm{d}s^2_{\text{AdS}_2}-\frac{1}{2u^2}\tilde{\theta}^1 \tilde{\theta}^2\mathrm{d}\tilde{\theta}^1 \mathrm{d}\tilde{\theta}^2\nonumber\\
&-\frac{i}{4u^2} \; \mathrm{d}u (\mathrm{d}\tilde{\theta}^1 \tilde{\theta}^2+\mathrm{d}\tilde{\theta}^2 \tilde{\theta}^1)+\frac{i}{4u^2}\; \mathrm{d}t (\mathrm{d}\tilde{\theta}^1 \tilde{\theta}^1+\mathrm{d}\tilde{\theta}^2 \tilde{\theta}^2),\\ 
\notag \\
\label{eq:B2Verlinde}
B_{\text{sAdS}_2}=&\frac{1}{4}(1 -\frac{2i}{u} \; \tilde{\theta}^1 \tilde{\theta}^2){\rm vol}_{\text{AdS}_2} - \frac{1}{4u^2} \tilde{\theta}^1 \tilde{\theta}^2 \left( \mathrm{d}\tilde{\theta}^1 \wedge \mathrm{d}\tilde{\theta}^1 - \mathrm{d}\tilde{\theta}^2 \wedge \mathrm{d}\tilde{\theta}^2 \right)\nonumber\\
&-\frac{i}{4u^2} \; \mathrm{d}u\wedge  (\mathrm{d}\tilde{\theta}^1 \tilde{\theta}^1+\mathrm{d}\tilde{\theta}^2 \tilde{\theta}^2)+\frac{i}{4u^2}\; \mathrm{d}t \wedge (\mathrm{d}\tilde{\theta}^1 \tilde{\theta}^2+\mathrm{d}\tilde{\theta}^2 \tilde{\theta}^1) \, , 
\end{align}
with $\mathrm{d}s^2_{\rm AdS_2}$ and vol$_{\text{AdS}_2}$ given in \eqref{warp}. 
By construction they are invariant under global OSp$(1|2)$ transformations.

\subsection{Extracting sAdS$_2$ from the dual model}
\label{subsec:Extracting sAdS$_2$ from the dual model}

The identification of the sAdS$_2$ factor \eqref{eq:dsVerlinde} inside the 3D metric \eqref{eq:PCDMPoincaremetric} is not straightforward. In fact, though the two metrics are written in terms of the same set of AdS$_2$ 
bosonic coordinates, the fermionic completions look very different. We are then forced to search for a map 
$(\theta^1, \theta^2)_{\eqref{eq:PCDMPoincaremetric}}  \leftrightarrow (\tilde\theta^1, \tilde\theta^2)_{\eqref{eq:dsVerlinde} }$ which allows to reconstruct the spinorial part of sAdS$_2$ in our dual metric \eqref{eq:PCDMPoincaremetric}. 
As long as this map is (locally) invertible we have the freedom to express the $\theta$'s in \eqref{eq:PCDMPoincaremetric} in terms of $\tilde\theta$'s or, viceversa, the $\tilde\theta$'s 
in \eqref{eq:dsVerlinde} in terms of $\theta$'s. For technical convenience, we prefer the second option. Therefore, we proceed as follows.

First, in \eqref{eq:dsVerlinde} we perform 
the following general coordinate transformation 
\begin{align}
\label{eq:tildethetaTothetaGeneral}
&\tilde{\theta}^1=c_{11}(t,u)\, \theta^1+c_{12}(t,u)\, \theta^2,\qquad\qquad \tilde{\theta}^2=c_{21}(t,u)\, \theta^1+c_{22}(t,u)\, \theta^2 \, ,
\end{align}
where $(\theta^1, \theta^2)$ are the spinorial coordinates used in \eqref{eq:PCDMPoincaremetric} and $c_{ij}$ are four arbitrary smooth functions on the two-dimensional $(t,u)$ slice. The explicit expression of the sAdS$_2$ metric in this new set of coordinates is given in eq. \eqref{eq:uplift}. Then, comparing with \eqref{eq:PCDMPoincaremetric}, 
we fix the  $c_{ij}$ unknowns in such a way that the OSp$(1|2)$ dual supermetric takes the form  
\begin{equation}
\label{eq:ansatz}
    \mathrm{d}s^2=  G_1(r,\theta) \; \mathrm{d}s^2_{\text{sAdS}_2}+G_2(r,\theta)\mathrm{d}r^2+G_3(r, \theta)\mathrm{d}(\theta^1\theta^2)\mathrm{d}r+G_4(r,\theta)\mathrm{d}\theta^1\mathrm{d}\theta^2 \, ,
\end{equation}
where $G_i$ are functions on the supermanifold, with expansion
\begin{equation}
\label{eq:Gexpansion}
    G_i(r,\theta) = g_i(r) + g_i^\theta(r)\, \theta^1\theta^2, \qquad i=1,2,3,4 \; .
\end{equation}

As detailed in appendix \ref{sec:matching}, solving the system of differential and algebraic equations which arise from matching the dual model \eqref{eq:PCDMPoincaremetric} with (\ref{eq:ansatz}, \ref{eq:uplift})  we find two possible solutions for $c_{ij}$. Since both describe the same supergeometry, without loosing generality we can focus on one of them. A brief discussion of the second solution can be found in appendix \ref{sec:matching}. 

The solution we consider here is \eqref{eq:hsSols}, which corresponds to the  following mapping 
\begin{align}
\label{eq:tildethetaTothetaExplicit}
\tilde{\theta}^1= t\theta^1+\theta^2 \, ,\quad\qquad \tilde{\theta}^2= u \, \theta^1 \, . 
\end{align}
Plugging it into eq. \eqref{eq:dsVerlinde} the  sAdS$_2$ metric reads,
\begin{equation}
    \begin{split}
    \label{eq:dsVerlindesol} \mathrm{d}s^2_{\text{sAdS}_2}=&\frac{1}{8}\mathrm{d}s^2_{\text{AdS}_2}-\frac{1}{2}\theta^1 \theta^2\mathrm{d}\theta^1 \mathrm{d}\theta^2-\frac{i}{4}\left(\mathrm{d}u \, \partial_t \mathcal{A} +\mathrm{d}t \, \partial_u \mathcal{A} \right) \, , 
    \end{split}
\end{equation}
where we used the definitions in \eqref{eq:Aderivatives}.

With $G$ functions in \eqref{eq:ansatz} correspondingly given by (note that $g_3^\theta$ is left unfixed)
\begin{equation}
\begin{alignedat}{3}  &G_1(r,\theta)&&=\frac{2r^2}{r^2-1}- i \frac{4r^2(2r^2+1)}{3(r^2-1)^2} \theta^1\theta^2 \, , &&\qquad \quad
G_3(r,\theta)=-\frac{ir(r^2-8)}{6(r^2-4)} + g_3^\theta (r) \theta^1\theta^2 \, , \\ &G_2(r,\theta)&&=\frac{1}{4}-\frac{i(r^4-4r^2-16)}{6(r^2-4)^2} \theta^1\theta^2 \, , &&\qquad \quad
    G_4(r,\theta)= \frac{2i}{3}+\frac{r^2+12}{18} \theta^1\theta^2 \, ,
\end{alignedat}
\end{equation}  
the dual supermetric \eqref{eq:PCDMPoincaremetric} finally is
\begin{align}\label{eq:sol2}
&\mathrm{d}s^2=\frac{2r^2}{r^2-1}\!\left(1-i \frac{2(2r^2+1)}{3(r^2-1)}\theta^1\theta^2\right)\mathrm{d}s^2_{\text{sAdS}_2}
\\
&+\!\frac{1}{4}\!\left(\!1\!-\!i\frac{2(r^4-4r^2-16)}{3(r^2\!-\!4)^2}\theta^1\theta^2\!\right)\!\mathrm{d}r^2\!-\!i\frac{r(r^2-8)}{6(r^2\!-\!4)}\mathrm{d}(\theta^1\theta^2)\mathrm{d}r\!+\frac{2i}{3}\!\left( \!1-\frac{i(r^2+12)}{12} \theta^1\theta^2 \! \right)\!\mathrm{d}\theta^1\mathrm{d}\theta^2\! .
\notag
\end{align}
The second row of this expression represents the line element of a $(1|2)$-dimensional supermanifold ${\mathcal M}^{(1|2)}$, a superline. The purely bosonic component of its metric is simply $g_{rr}|_{\theta=0}= 1/4$, whereas it exhibits a $r$-dependent non-trivial geometry in the fermionic sector.

We stress that, while sAdS$_2$ is invariant under OSp$(1|2)$, in particular under the supersymmetry transformations generated by the spinorial charges of the $\mathfrak{osp}(1|2)$  algebra, ${\mathcal M}^{(1|2)}$ is not invariant. Rather, its supersymmetry transformations have to compensate for the non-trivial transformation of the warp factor in front of $\mathrm{d}s^2_{\text{sAdS}_2}$, in order to make the three-dimensional $\mathrm{d}s^2$ invariant. The only isometry exhibited by the superline is a SL$(2, {\mathbb R})$ symmetry which rotates the two spinorial coordinates, leaving $r$ untouched. In sAdS$_2$ this symmetry is violated by the $\partial_t \mathcal{A}, \partial_u \mathcal{A}$ terms in \eqref{eq:dsVerlindesol}.

Using the same procedure,  we can rewrite the $B$ field of the dual model such that a $B_{\text{sAdS}_2}$ factor appears. To this end, we first apply transformation \eqref{eq:tildethetaTothetaExplicit} to the $B$ field of the sAdS$_2$ background, equation \eqref{eq:B2Verlinde}, obtaining the simpler expression 
\begin{align}
\label{eq:B2Verlindesol2}
B_{\text{sAdS}_2}=&-\frac{1}{4}\left({\rm vol}_{\text{AdS}_2}+\theta^1\theta^2 \mathrm{d}_F\mathcal{A}-i\mathrm{d}_B\mathcal{A}\right),
\end{align}
then, comparing this expression with the $B$ field for the dual model given in \eqref{eq:PCDMPoincareB2}, we can easily realize that the latter can be recast into the following form
\begin{equation}
\begin{aligned}
\label{eq:sol2B2}
    B\!=&\!-\!\frac{2r^3}{r^2\!-\!1}\left(\!1\!-\!\frac{i(r^2\!+\!5)}{3(r^2\!-\!1)}\theta^1\theta^2\!\right)B_{\text{sAdS}_2}
    \\
    &+\frac{ir}{3}\left(1\!-\!\frac{10i}{3}\theta^1\theta^2\!\right)\left(\!\mathrm{d}_B\mathcal{A}\!-\!\frac{1}{2}\mathrm{d}_F\mathcal{A}\!\right)
    -\frac{4i}{3(r^2-4)}\mathrm{d}r\wedge\mathcal{A} .
\end{aligned}
\end{equation}

\vspace{10pt}
We managed to rewrite the metric and the $B$ field of the OSp$(1|2)$ dual model in a convenient set of supercoordinates where a sAdS$_2$ component is explicitly present, together with a $(1|2)$-dimensional completion. Though this may sound like the supersymmetric analogue of the factorization \eqref{warp} occurring in the bosonic dual model, there is a profound difference between the two cases. In the bosonic case, the AdS$_2$ metric depends only on the two Poincar\'e coordinates, as it should, and the third coordinate $r$ is factorized out in a warp factor. This leads to an actual factorization of the geometry into AdS$_2 \times \mathbb R^+$. In the supersymmetric case, instead, the complete factorization is prevented by supersymmetry, which requires both sAdS$_2$ and ${\mathcal M}^{(1|2)}$ to depend on the same set of spinorial coordinates. 

Another interesting observation concerning \eqref{eq:sol2} is that the $r^2 = 4$ singularity is completely encoded into the superline, while the warp factor in front of $\mathrm{d}s^2_{\text{sAdS}_2}$ exhibits only the physical singularity at $r^2=1$. This is a 
further confirmation of the non-physical nature of the $r^2 = 4$ singularity. In principle, this could be eliminated by an appropriate $r$-dependent rescaling of the T-dual $\theta$ variables. However, this would come at the price of introducing a parametric $r$ dependence in the sAdS$_2$ sector.

\vskip 15pt

To complete the picture, we now apply the same procedure to identify a sAdS$_2$ factor inside the dual geometry \eqref{eq:metric-poincare-sl2-natd}, obtained by dualising the maximal bosonic subgroup SL$(2,\mathbb R)_{L}$ contained in the left sector of the isometry group. 

Proceeding as in the previous case, we first perform the change of spinorial variables \eqref{eq:tildethetaTothetaGeneral} to write the sAdS$_2$ metric as in \eqref{eq:uplift}.  We then exploit the arbitrariness of the $c_{ij}$ functions driving the change of coordinates to bring the dual supermetric, eq.  \eqref{eq:metric-poincare-sl2-natd}, to the form \eqref{eq:ansatz}. This leads to a system of differential and algebraic equations for  $c_{ij}$, and the functions $g_i, g_i^\theta$ in \eqref{eq:Gexpansion}, which is exactly the same as the one of the previous section. As detailed in appendix \ref{sec:matching}, we find two  solutions for $c_{ij}$, both describing the same supergeometry. Therefore, without loosing generality we can focus on the one leading  to the following mapping 
\begin{align}
\label{eq:bosonictildethetaTothetaExplicit}
\tilde{\theta}^1=\;t\theta^1+\theta^2,\quad\qquad \tilde{\theta}^2=\;u\theta^1.
\end{align} 
Correspondingly, after rescaling the fermions in the T-dual model \eqref{eq:metric-poincare-sl2-natd} by $\theta^{\alpha} \rightarrow i\sqrt{2} \theta^\alpha$ in analogy with \eqref{susy-dualisation-fermionic-rescaling}, the functions defined in \eqref{eq:ansatz}  turn out to be given by 
\begin{equation}
G_1(r,\theta)=\frac{2r^2}{r^2-1}, \quad\,\,\,   G_2(r,\theta)=\frac{1}{4},  \quad\,\,\, G_3(r,\theta)=g_3^{\theta}\theta^1\theta^2,  \quad\,\,\,
G_4(r,\theta)=2(i +  \theta^1 \theta^2) \, ,
\end{equation}
with $g_3^{\theta}$ remaining an arbitrary function.

The final form of the dual metric with a sAdS$_2$ factor explicitly factorized then reads 
\begin{align}
\label{eq:bosonicsol2}
\mathrm{d}&s^2=\frac{2r^2}{r^2-1}\mathrm{d}s^2_{\text{sAdS}_2}+\frac{1}{4} \mathrm{d}r^2+2i\bigl(1-i\theta^1\theta^2\bigr)\mathrm{d}\theta^1\mathrm{d}\theta^2\, , 
\end{align}
with the $\mathrm{d}s^2_{\text{sAdS}_2}$ slice given by the expression \eqref{eq:dsVerlindesol}. Similarly, upon transformation \eqref{eq:bosonictildethetaTothetaExplicit}, for the $B$ field we obtain 
\begin{align}
\label{eq:sol2B2-2}
B=&-\frac{2r^3}{r^2-1}B_{\text{sAdS}_2}+\mathrm{d}\left[\frac{ir\mathcal{A}}{2}\right]-\frac{ir}{2}(1-i\theta^1\theta^2)\mathrm{d}_{F}\mathcal{A}\, , 
%B=&-\frac{2r^3}{r^2-1}B_{\text{sAdS}_2,\theta} -\frac{r(2r^2+1)}{4(r^2-1)}\theta^1\theta^2\mathrm{d}_{F}\mathcal{A}+\frac{ir}{2}\mathrm{d}_{B}\mathcal{A} +\frac{i}{2} \mathrm{d}r \wedge \mathrm{d}\mathcal{A} \, , 
\end{align}
where the $B$ field in the sAdS$_2$ background is given by \eqref{eq:B2Verlindesol2}.

We note that the supermetric in \eqref{eq:bosonicsol2} describes a factorized sAdS$_2 \times {\mathcal M}^{(1|2)}$ geometry, with the line element $\mathrm{d}r^2$ being the same as the one of the SL$(2,\mathbb{R})$ model.  This is also related to the fact that, in contrast with the case of dualisation of the full OSp$(1|2)_{L}$ sector, now mixed terms $\mathrm{d}(\theta^1\theta^2)\mathrm{d}r$ do not appear. 
Indeed, when dualisation is carried out on the SL$(2,\mathbb{R})_{L}$ bosonic subgroup, the fermionic sector remains as a spectator and, in particular, it does not interfere with the bosonic sector, thus leaving $\mathrm{d}r$ untouched.

\section{(Super)NATD and JT (super)gravity}\label{sec:JT}

From the previous discussion, we can conclude that applying (S)NATD to PCM models describing (super)AdS$_3$ geometries leads to a dual geometry that roughly speaking is of the form
\begin{equation}
\label{eq:reduction}
\begin{aligned}
& {\rm AdS}_3 \; \longrightarrow \; {\rm AdS}_2 \, \! \! \times {\rm line}  \\
& {\rm sAdS}_3 \longrightarrow {\rm sAdS}_2 \, \! \! \times {\rm superline} 
\end{aligned}    
\end{equation}
On the other hand, (s)AdS$_2$ geometries are described by 2D dilaton (super)gravities, and enjoy a nice description in terms of 3D Poisson sigma models where one of the three directions is identified with the dilaton field. It is therefore tempting to interpret the reduction in \eqref{eq:reduction} in this perspective, with the third direction playing  the role of the dilaton (plus the dilatino, in the supersymmetric case). As we are going to discuss in the rest of this section, such an identification is not straightforward, however we find that introducing a suitable coupling constant in the original PCM, in the small coupling regime the dual action gives rise to Jackiw-Teitelboim (JT) gravity \cite{Jackiw:1984je, Teitelboim:1983ux} and its ${\cal N}=1$ supersymmetric generalization \cite{Teitelboim:1983uy,Chamseddine:1991fg,Izquierdo:1998hg,Ikeda:1993fh,Ikeda:1993dr}. Since this requires exploiting the connection between JT gravity and Poisson sigma models, first we briefly review this connection. 

\subsection{(Graded) Poisson sigma models and JT (super)gravity: a brief review}

It is well known that dilaton gravity actions in 2D can be written as 3D Poisson sigma models where one of the target space coordinates is identified with the 2D dilaton field. In the bosonic case this was originally proposed in \cite{Schaller:1994es, Schaller:1994pm} 
 (for a nice review, see \cite{Mertens:2022irh}), whereas the generalization to 2D supergravity was worked out in \cite{Strobl:1999zz, Ertl:2000si, Bergamin:2002ju}. Here we briefly review both cases. 

A bosonic sigma model on a two-dimensional worldsheet $\Sigma$ is said to be of the Poisson type if its background manifold $\mathcal{M}$ is Poisson. This means there exists an antisymmetric and bilinear map $P$, known as Poisson bracket, acting on smooth functions on $\mathcal{M}$, and satisfying the Leibnitz rule and the Jacobi identity (we refer to 
\cite{Schaller:1995xk}
for a useful review of Poisson sigma models).
Given a set of local coordinates $X^{a}$ on $\mathcal{M}$ (smooth maps from the worldsheet $\Sigma$ to the target space), it is sufficient to define the elementary Poisson structure $P^{ab}(X)\equiv \{X^{a},X^{b} \}$. The bracket of any other pair of functions $f,g$ can be then expressed in terms of such a tensor as
$\{f,g\} = P^{ab}(X)\, \partial_{a} f \, \partial_{b}g$, 
where $\partial_{a}f \equiv \tfrac{\partial f}{\partial X^{a}}$. 

Introducing a set of 1-form fields on $\Sigma$, $A_a = d\sigma^i A_{ia}$ with $a\in \{0,1,2\} , \, i\in \{0,1\}$, the action of a Poisson sigma model is defined as
\begin{equation}\label{Poisson-action0}
S_{PSM}= \int_{\Sigma} \left(  \textrm{d}X^{a} \wedge A_a + \frac{1}{2} P^{ab}(X)A_b\wedge A_a  \right) + \int_{\Sigma} \text{Vol}(\Sigma) C(X) \, \equiv \,S_{PSM}^{Top} + S_{PSM}^{C} \,\, .
\end{equation}
 Lacking any reference to the worldsheet metric, the first integral represents the topological part of the Poisson sigma model. Instead the second part, which depends on the Casimir function $C(X)$ for the Poisson structure spoils this property due to the presence of the volume form on $\Sigma$. 

The topological part of the action in \eqref{Poisson-action0} can be shown to be locally equivalent to JT gravity \cite{Jackiw:1984je,Teitelboim:1983ux} when we restrict to a three-dimensional Poisson manifold and choose a Poisson structure which is linear in the $(X^0, X^1, X^2)$ coordinates, 
$P^{ab}(X) \equiv -2X^{c}\varepsilon_{c}{}^{ab}$, with $\varepsilon_{abc}$ being the antisymmetric Levi-Civita tensor in 3D\footnote{More general dilaton gravity actions can be constructed, which correspond to choosing non-linear Poisson structures (for a general discussion, see \cite{Ertl:2000si}).}.
In fact, identifying the $X^2$ coordinate with a scalar field $\Phi$ and renaming $A_a = (-\tfrac{1}{2}e_i, -\tfrac{1}{2}\omega)$, with $i\in \{0,1\}$, 
the topological Poisson action reduces to\footnote{The lowered-index one-forms and Poisson structure have been conveniently chosen, so as to match our conventions in appendix \ref{appendix:osp-algebra}.} 
\begin{equation}
S_{PSM}^{Top} = 
\frac{1}{2}\int_{\Sigma} \left[ \Phi \textrm{d}\omega - \frac{1}{2}\Phi \varepsilon^{ij}e_{j}\wedge e_{i} +X^{i}(\textrm{d}e_{i}-\varepsilon_{i}{}^{j}\omega \wedge e_{j}) \right]
\, \, , 
\end{equation}
where we defined the 2D Levi-Civita symbol as $\varepsilon_{ij} \equiv \varepsilon_{ij2}$. Now, integrating out $X^i$ enforces the torsion-free constraints $(\textrm{d}e_{i}-\varepsilon_{i}{}^{j}\omega \wedge e_{j})=0$, which allow to interpret $\omega^{ij} = \varepsilon^{ij} \omega$ as the spin connection of a 2D geometry described by the $e_i$ zweibein. As a consequence, we can rewrite $e_0 \wedge e_1 = \mathrm{d}^2x \sqrt{-g}$ and $2 \mathrm{d}\omega = \mathrm{d}^2x \sqrt{-g} R$, where $g$ is the determinant of the 2D metric and $R$ the corresponding scalar curvature. The action thus reduces to
\begin{equation}
\label{eq:JT}
S_{PSM}^{Top} =
\frac{1}{4} \int_{\Sigma} \mathrm{d}^2x \sqrt{-g} \, \Phi (
R + 2) \, ,
\end{equation}
which is exactly the JT action describing dilaton gravity on AdS$_2$ geometry with cosmological constant $\Lambda = -2$, and no boundary \cite{Jackiw:1984je,Teitelboim:1983ux} . 

\vskip 10pt

The definition of Poisson sigma model  can be generalized to graded Poisson sigma model \cite{Strobl:1999zz}, which includes supersymmetry among the set of its symmetries. 
Specializing to ${\cal N}=1$ graded models in three dimensions, 
the defining action is given by \cite{Strobl:1999zz, Ertl:2000si}
\begin{equation}\label{Graded-Poisson-action0}
S_{GPSM}= \int_{\Sigma} \left(  \textrm{d}X^{A}\wedge A_{A} + \frac{1}{2}P^{AB}(X) A_{B}\wedge A_{A} \right) + \int_{\Sigma} \text{Vol}(\Sigma) C(X) \, \equiv \,S_{GPSM}^{Top} + S_{GPSM}^{C}
\end{equation}
where $X^A=(X^a,\theta^\alpha), \, a \in \{ 0,1,2\}, \, \alpha \in\{1,2\}$, is a set of graded coordinates in ${\cal N}=1$ superspace, $A_A = (A_a, A_\alpha)$ are graded one-forms on the worldsheet $\Sigma$, whereas $P^{AB}$ is a graded antisymmetric Poisson structure, $P^{AB} = -(-1)^{AB} P^{BA}$, satisfying the graded version of Jacoby identities. Finally, $C(X)$ is a Casimir of the Poisson superalgebra.  

Along the lines reviewed above, graded Poisson sigma models can be exploited to obtain dilaton supergravity actions in two dimensions. In this case we set $X^A \equiv (X^i, \Phi, \theta^\alpha)$, with $i \in \{ 0,1\}$, still identifying one of the bosonic coordinates with the dilaton field, $\theta^\alpha$ with the dilatino components, and $A_A \equiv (-\tfrac{1}{2}e_i, -\tfrac{1}{2}\omega, -i\psi_\alpha)$ providing the zweibein, the spin connection and the Rarita-Schwinger one-form of the 2D supergravity. 

Concerning the Poisson structure, compatibility with target space Lorentz invariance fixes the mixed components to be  $P^{i\Phi} = -2X^j \varepsilon_j^{\; \, i}$,\; $i,j \in \{0,1\}$, and  $P^{\alpha \Phi} =  \theta^\beta (\gamma^3)_\beta^{\; \; \alpha}$ \footnote{Here $\gamma^3 \equiv \gamma^0 \gamma^1$ is the 2D analogue of $\gamma^5$ in four dimensions. Consistently with the dimensional reduction, it coincides with $\gamma^2$ of the 3D Clifford algebra (see eq.\eqref{explicit-representation-gamma-matrices}).}. Therefore, if we define $e_A \equiv (-\tfrac{1}{2}e_i, -i\psi_\alpha)$ the topological part of \eqref{Graded-Poisson-action0} can be rewritten as 
\begin{equation}
S^{Top}_{GPSM} \! = \! \frac{1}{2} \! \int_{\Sigma} \! \left[ \! \Phi \textrm{d}\omega 
\! + \! X^{i}(\textrm{d}e_{i} \! - \! \varepsilon_{i}{}^{j}\omega \wedge e_{j}) 
\! + \! 2i\theta^\alpha \left( \! \mathrm{d}\psi_\alpha \! + \! \frac12 (\gamma^3)_{\alpha}{}^{\beta}\omega \wedge \psi_\beta \! \right) \! + \!  P^{AB}(X)  e_B \wedge e_A
 \! \right], \! 
\end{equation}
According to the choice of $P^{ij}, P^{i\alpha}$ and $P^{\alpha\beta}$ appearing in the last term, we  obtain the action for different dilaton supergravities \cite{Cangemi:1993mj,
Ikeda:1993dr,Ikeda:1993fh,Izquierdo:1998hg,Strobl:1999zz,Ertl:2000si}. For our purposes, here we are interested in the case where $P^{\alpha i}= \theta^\beta (\gamma^i)_{\beta}{}^{\alpha}$,\; $P^{ij} = -2\Phi \varepsilon^{ij}$ and $P^{\alpha\beta} = -\tfrac{i}{2}\Phi (\gamma^3)^{\alpha\beta} -\tfrac{i}{2} X_i(\gamma^i)^{\alpha\beta}$. 
With this choice the previous action becomes
\begin{align}
S^{Top}_{GPSM} & = \frac{1}{2} \int_{\Sigma} \Bigg[ \Phi \textrm{d}\omega 
- \frac{1}{2}\Phi \, \varepsilon^{ij} \, e_j \wedge e_i + X^{i}\left( \textrm{d}e_{i}-\varepsilon_{i}{}^{j}\omega \wedge e_{j} +\frac{i}{2} (\gamma_{i})^{\alpha\beta} \psi_\beta \wedge \psi_\alpha \right) 
\\
& \qquad \quad \,\, + \frac{i}{2}  \Phi (\gamma^3)^{\alpha\beta} \psi_\beta \wedge \psi_\alpha  + 2i \theta^\alpha \left( \mathrm{d}\psi_\alpha \! + \! \frac12 (\gamma^3)_{\alpha}{}^{\beta}\omega \wedge \psi_\beta \! +\! \frac12 (\gamma^i)_{\alpha}{}^{\beta} e_i \wedge \psi_\beta \right)  \Bigg]  .
\notag
\end{align}
Now, the equations of motion for $X^i$ are exactly the standard supertorsion constraints of 2D superspace, whose solution $\omega$ is then identified with the spin connection. Therefore, proceeding as in the bosonic case we finally obtain 
\begin{align}\label{eq:JTSUGRA}
S^{Top}_{GPSM} & = \frac{1}{4}  \int_{\Sigma} \mathrm{d}^{2}x  \Bigg[ \Phi  \! \left( \sqrt{-g}(R +2) + i\varepsilon^{kl}(\gamma^{3})^{\alpha\beta}\psi_{k\beta} \psi_{l\alpha}  \right) 
\\
& \qquad \qquad \qquad \quad +4i\epsilon^{kl}\theta^\alpha \left( \partial_{k}\psi_{l\alpha} + \frac{1}{2} (\gamma^3)_{\alpha}{}^{\beta}\omega_{k}  \psi_{l\beta} +\frac{1}{2} (\gamma^i)_{\alpha}{}^{\beta} e_{ki} \psi_{l\beta} \right)  \Bigg] \,\, .
\notag
\end{align}
This is the supersymmetric generalization of the JT action \eqref{eq:JT}, first proposed in \cite{Chamseddine:1991fg}. It describes a sAdS$_2$ background with cosmological constant $\Lambda = -2$. 

In the rest of this section we are going to show that in a suitable limit the action for the SNATD of the OSp$(1|2)$ PCM reduces exactly to \eqref{eq:JTSUGRA}. In particular, this implies that setting the spinorial coordinates and the gravitino to zero the NATD model of the bosonic SL$(2, \mathbb{R})$ PCM reduces to the JT gravity action in \eqref{eq:JT}.

\subsection{JT (super)gravity from (super)NATD}\label{subsec:JT_from_NATD}

In order to introduce our procedure and clarify which is the limit where the (super)AdS$_2$ geometry should arise, we warm up by considering the bosonic PCM on SL$(2, \mathbb{R})$ before dualisation. It describes AdS$_3$ geometry with a vanishing $B$ field. 

Choosing for convenience global coordinates $Z^a = (t,u,r)$, we write the AdS$_3$ 
metric as a 2D metric with  $(t,u)$ coordinates describing AdS$_2$ geometry plus a warp function of $u$ times the ${\textrm d}r$ line element,
\begin{equation}
\label{eq:AdS3metric}
   \mathrm{d} s^2_{\rm{AdS}_3} = \frac{-\mathrm{d}t^2 + \mathrm{d}u^2 + \mathrm{d}r^2}{u^2} = \mathrm{d}s^2_{{\rm AdS}_2} + \frac{1}{u^2}\mathrm{d}r^2 \, . 
\end{equation}

Although the SL$(2, \mathbb{R})$ PCM has vanishing $B$ field, it is convenient to exploit gauge invariance to add a pure gauge $\hat B = \mathrm{d} C$. The sigma model action then reads (we stick to the notations of section \ref{sec:review})
\begin{equation}\label{eq:originalbosonic}
S= -\frac{\lambda}{2}  \int_{\Sigma} \mathrm{d}Z^a \wedge (1+\star)\, \mathrm{d}Z^b \, (\hat{G} + \hat{B})_{ba} \, , 
\end{equation} 
where we have introduced a suitable coupling constant $\lambda$. The metric $\hat G_{ab}$ can be read in \eqref{eq:AdS3metric}, whereas  we fix $\hat B$ as \begin{equation}
    \begin{split}
        \hat B=\frac{2}{u^2}(\mathcal{X} \, \mathrm{d}t\wedge \mathrm{d}r+\mathcal{Y} \, \mathrm{d}u\wedge \mathrm{d}r +\mathcal{Z} \, \mathrm{d}t\wedge \mathrm{d}u),
    \end{split}
\end{equation}
with $\mathcal{X}$, $\mathcal{Y}$ and $\mathcal{Z}$ being smooth functions of the $(t,u,r)$ coordinates and the coupling $\lambda$. In order to ensure that $\partial_{[a} \hat B_{bc]} =0$, they  have to satisfy the following differential equation
\begin{equation}
 \label{eq:puregaugeB-bosonic0}
        \partial_u \left(\frac{\mathcal{X}}{u^2}\right)-  \partial_t \left(\frac{\mathcal{Y}}{u^2}\right)-\partial_r \left(\frac{\mathcal{Z}}{u^2}\right)=0\, .
\end{equation}

We now rewrite the action in the first order formalism. Introducing one forms $A_a$ on $\Sigma$, it is easy to show that  \eqref{eq:originalbosonic} is equivalent to 
\begin{equation}
\label{eq:firstorderaction}
S = -\int_\Sigma \left[  A_a \wedge \mathrm{d} Z^a  + \frac{1}{2\lambda} \, \left(  \hat g \,  + \, \lambda \hat P \right)^{\! ab} \, A_b \wedge (1+\star)\, A_a   \right] \,\, ,
\end{equation}
where we have defined 
\begin{equation}
\label{eq:Phat}
    \begin{split}
        (\hat G \!  + \! \hat B)^{-1} \equiv \hat g \! \, + \, \lambda \hat P \, , \qquad {\rm with} \qquad  \lambda \hat P=\frac{ u^2}{(\mathcal{X}^2+\mathcal{Z}^2-\mathcal{Y}^2-1)}\begin{pmatrix}
0 & -\mathcal{Z} & -\mathcal{X}
\\
\mathcal{Z} & 0 & \mathcal{Y} 
\\
\mathcal{X} & -\mathcal{Y} & 0
\end{pmatrix} \, \, ,
    \end{split}
\end{equation}
and $\hat{g}$ given by the symmetric part of $(\hat G \! + \! \hat B)^{-1}$ .
It is easy to see that replacing the $A_a$ forms in \eqref{eq:firstorderaction} with their equations of motion we are back to \eqref{eq:originalbosonic}.

Now, the crucial point is that taking the $\lambda \to \infty$ limit, while keeping $\hat g$ and $\hat P$ fixed\footnote{This is similar in spirit to the $\alpha'\to 0$ Seiberg-Witten limit \cite{Seiberg:1999vs} under which string theory gives rise to spacetime non-commutativity. It has been used in \cite{Baulieu:2001fi} to reduce the closed string sigma model to a Poisson sigma model.}, we get rid of the $\hat g$ term in \eqref{eq:firstorderaction} thus landing on a Poisson-like action with $\hat P$ playing the role of the Poisson structure. 
In this limit, we expand the unknown functions in powers of $\lambda^{-1}$ and solve differential equation \eqref{eq:puregaugeB-bosonic0} and Jacobi identities for $\hat P^{ab}$ order by order in $\lambda^{-1}$. A detailed inspection reveals that if we choose the order zero components of $\hat P$ to be linear in the coordinates, precisely $\hat P^{02} = -c_1 u$, $\hat P^{12} = -c_1 t$ and $\hat P^{01} = c_1 r+c_2$, with $c_1, c_2$ constants, the pure gauge condition \eqref{eq:puregaugeB-bosonic0} for $\hat B$ is identically satisfied and the $\hat P$ structure satisfies Jacobi identities, up to ${\cal O}(\lambda^{-1})$ terms that can be safely ignore for $\lambda \to \infty$. 

Without loosing generality we can choose $c_1=-2$ and $c_2=0$, so that 
$\hat P^{ab}  = -2X^{c}\epsilon_{c}{}^{ab} $. 
Finally, if we identify the $r$ coordinate in \eqref{eq:AdS3metric} with the dilaton $\Phi$ and follow the reasoning summarised in the previous section, from the action in \eqref{eq:firstorderaction} we recover the JT gravity action \eqref{eq:JT} describing AdS$_2$ geometry.\\

A more interesting question is whether this procedure can be used to reduce the (S)NATD model to the JT (super)gravity action. As already mentioned, this would be somehow expected, as the dual metric always factorizes into a warped (super)AdS$_2$ part times a one-dimensional (super)manifold. 

For simplicity, we first discuss the bosonic SL$(2, \mathbb{R})$ model, whose dual geometry clearly factorizes as in \eqref{warp} when we apply the coordinate transformation \eqref{eq:paramet}. In the convenient $\{t,u,r\}$ coordinates it would be natural to identify $r$ with the dilaton and apply the reduction described above to dig out the JT action. However, while these coordinates make the AdS$_2$ factor manifest in the dual model, the above limiting procedure becomes quite cumbersome and of no simple interpretation. Therefore, we find more convenient to work directly in the set of dual coordinates coming from the Lagrange multipliers.

Given the original PCM \eqref{eq:originalbosonic} with $\hat B = 0$,
the NATD procedure gives rise to an action of the form \eqref{T-dual-sigma model-paper-notation}
\begin{align}\label{T-dual-sigma model-paper-notation2}
\tilde S &= -\frac{1}{2\lambda} \int_{\Sigma} \,\, \langle \textrm{d} X, \frac{1}{1-\lambda^{-1}\textrm{ad}_{X}} (1+ \star) \textrm{d}X \rangle \qquad \text{with} \qquad 
X \equiv \texttt{g}^{-1}\Lambda \texttt{g} \equiv X^a T_a  \, , \notag \\
& = -\frac{1}{2\lambda} \int_{\Sigma} \mathrm{d}X^a \wedge (1+\star)\, \mathrm{d}X^b \, (G + B)_{ba} \, , 
\end{align}
where $T_a$ are the $\mathfrak{sl}(2, \mathbb{R})$ generators, and along dualisation we kept track of the $\lambda$ coupling introduced in \eqref{eq:originalbosonic}. Having dualised the full left sector of the isometry group, we also have the freedom to choose the gauge $\texttt{g}=\mathbb{1}$, so that $X$ only contains the multipliers.
As discussed in appendix \ref{appendix:osp-algebra} the dual metric and $B$ field are respectively identified as the symmetric and antisymmetric components of the operator $\frac{1}{1-\lambda^{-1}\mathrm{ad}_X}=\sum_{k=0}^\infty \lambda^{-k}\mathrm{ad}_X^k$.
Therefore, setting
\begin{equation}
\label{eq:gPdefinition}
(G+B)^{-1} = 1-\lambda^{-1}\mathrm{ad}_X\equiv  g - \frac{1}{\lambda} P \, , \qquad {\rm with} \qquad g = 1 \qquad \text{and} \qquad 
P =  \, \mathrm{ad}_X \, ,
\end{equation}
and introducing 1-forms $A\equiv A^aT_a$ on the worldsheet, we can recast the dual action in the  first order formalism as
\begin{equation}
\label{eq:firstorder-index-free}
\tilde S = -\int_{\Sigma} \langle  A,\mathrm{d}X \rangle + \frac{\lambda}{2} \langle A, \star  A \rangle -  \frac{1}{2}\langle  A,  P(A) \rangle \,\, ,
\end{equation}
In the $\lambda \to 0$ limit,  the second term vanishes and we are left with a Poisson sigma model, which in components reads
\begin{align}
\tilde S& \underset{\lambda \to 0}{=} \int_{\Sigma} \left[ \textrm{d}X^{a} \wedge A_a + \frac{1}{2}P^{ab}(X) A_{b}\wedge A_{a} \right],
\end{align}
with the Poisson structure identified as $P^{ab}(X) = \delta^{ad}X^{c}f_{cd}{}^{b}$ (with $\delta^{ad} \equiv -2 \eta^{ad}$, see  \eqref{eq:delta}), and automatically satisfying the Jacobi identities. 
Now, renaming $X^{a}=(X^{i},X^{2}\equiv \Phi)$ and $A^{a}=(e^{i},\omega)$, with $i \in \{0,1\}$, it is easy to see that the above action reduces to the JT action for AdS$_2$ background, given in \eqref{eq:JT}. \\

Generalization to the supersymmetric case is straightforward, as the action dual to the PCM on OSp$(1|2)$ is again \eqref{T-dual-sigma model-paper-notation}, where the Lie algebra valued $X$ has to be replaced with the Lie superalgebra valued $X \equiv X^A T_A$, containing three bosonic and two fermionic components $X^A = (X^i,X^2,\theta^\alpha)$, with $i\in \{0,1\}, \, \alpha \in \{1,2\}$. We can then proceed as before, obtaining \eqref{eq:firstorder-index-free} with $A$ providing graded worldsheet 1-forms, $A^A = (A^i,A^2,A^\alpha)$.
If we now apply the $\lambda \to 0$ limit with $g$ and $P$ in \eqref{eq:gPdefinition} fixed, we land on
\begin{equation}\label{T-dual-Poisson-supersymmetric-case}
\tilde S \underset{\lambda \to 0}{=}   \int_{\Sigma} \left[ \textrm{d}X^{A} \wedge A_A + \frac{1}{2}P^{AB}(X) A_{B}\wedge A_{A} \right] \,\, .
\end{equation}
Now, identifying $X^2 = \Phi$ and $A^{A} = (e^i, \omega, \psi^\alpha)$, and taking into account that $P^{AB}(X) = \delta^{AD}X^{C}f_{CD}{}^{B}$ (with  $\delta^{AD}$ defined in \eqref{eq:delta}) is a graded Poisson structure with even and odd components, following the previous steps we eventually obtain the supersymmetric generalization of JT action given in \eqref{eq:JTSUGRA}. 

Summarizing, we have shown that in the $\lambda \to 0$ limit the T-dual action gives rise to the JT action for 2D (super)gravity, consistently with the reduction scheme \eqref{eq:reduction} observed under (S)NATD. 

We conclude by observing that, despite having applied the above limiting procedure to specific PCMs, this can in principle be applied to any PCM associated to a generic (super)group G. Here, we briefly comment on  some universal features underlying it.

First of all, we recall that applying the dualisation procedure to a generic PCM on G with a $\lambda$-coupling requires generalizing the master action in \eqref{master-action} as 
\begin{equation}\label{master-action-with-coupling}
S_{\omega} = -\tfrac{\lambda}{2} \int_{\Sigma} \langle j_\omega, \star j_\omega \rangle - \int_{\Sigma} \langle \Lambda, F_{\omega} \rangle \, .
\end{equation}

On the other hand, rewriting the T-dual action \eqref{T-dual-sigma model-paper-notation2} in the first order form \eqref{eq:firstorder-index-free}, it is not hard to realise that integrating by parts the first term and rearranging the last one, the action exhibits the same structure as the master action \eqref{master-action-with-coupling}, where the gauged Maurer-Cartan form $j_{\omega}$ is replaced by the one-form $A$, after using the identity $\langle \Lambda, F_\omega \rangle = \langle X, F_{j_{\omega}} \rangle$, with $X\equiv \texttt{g}^{-1}\Lambda \texttt{g}$.

In particular, if we gauge the full left (or right) sector of the G$_L \times$G$_R$ isometry group, as we have done above, the freedom to set $\texttt{g}=\mathbb{1}$ can be exploited to reduce $j_\omega \rightarrow \omega$, and the first order form of the T-dual action exactly matches the master action upon formally identifying $\omega \leftrightarrow A$. This explains why in the two cases discussed above, using the Lagrange multiplier coordinates allows for an easy identification with the JT action. The situation would become more involved if we were to gauge only a subgroup of the original isometry group. This more general case would deserve a deeper investigation which is beyond the scope of this paper.

Finally, it should be stressed that the whole mechanism is a result of the independence of the coefficients which appear in front of the two terms in the master action \eqref{master-action}: since each term is separately invariant under the gauged left (or right) sector of the isometry group, the two coefficients can be completely unrelated.

\vskip 10pt

\section{Discussion}\label{sec:discussion}

\begin{figure}[t]
\centering
\includegraphics[scale=0.8]{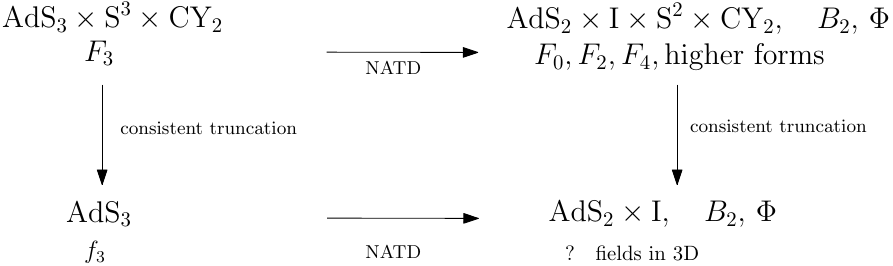}
\caption{Ideal commutating diagram between NATD and consistent truncation.}
\label{diagrama}
\end{figure}

In this paper we have shown how to construct super-AdS$_2$ backgrounds dualising super-AdS$_3$ spaces. Our construction is manifestly supersymmetric and consistently reduces to the bosonic findings of \cite{Ramirez:2021tkd, Ramirez:2022fkc} when we remove the spinorial coordinates and the fermion-fermion block of the supermatrix generators. 

We should stress that our original model, upon removing the fermionic variables and including a 3-form field strength $f_3$, can be obtained from a consistent truncation of the near horizon geometry of the D1-D5 brane system, along the lines of \cite{Cvetic:2000dm,Deger:2014ofa} (left side of figure \ref{diagrama}). This solution satisfies the equations of motion of a 3D gauged supergravity \cite{Cvetic:2000dm}. We have shown how the metric of the truncated theory transforms under NATD. It would be interesting to include $f_3$ in the dualisation process and study how it transforms. It is expected to give rise to a wider spectrum of forms and scalar fields (bottom of figure \ref{diagrama}).  
On the other hand, applying NATD directly to AdS$_3\times$S$^3\times$CY$_2$ one obtains the AdS$_2\times$I$\times$S$^3 \times$CY$_2$ background \cite{Ramirez:2021tkd} (upper of figure \ref{diagrama}).
It would be interesting to study the consistent truncation of this dual geometry to 3D (right side of figure \ref{diagrama}). One would expect this to match our solution given in \eqref{eq:sol2} and \eqref{eq:sol2B2}, with the addition of extra fields produced by T-dualising $f_3$. This would prove that the two operations (NATD and consistent truncations) commute. As a consequence, one might interpret the superconformal quantum mechanics studied in \cite{Ramirez:2021tkd} as the dual field theory of our solution. 
 
In section \ref{sec:JT} we applied a limiting procedure to reduce the 3D T-dual action to JT (super)gravity. Our findings are a special case of a more general scheme, which reveals a very close connection between NATD of PCMs and topological models. This might lead to interesting new results. For example, in the context of 10D string theory, where in many cases part of the compact geometry sector is described by PCMs, we can identify the original coupling $\lambda$ with the string tension $1/\alpha '$. In the strong coupling limit, $\lambda \to \infty$, the fields associated to the compact subgroup would enjoy a purely topological description \cite{Baulieu:2001fi}. On the other hand, for PCM sectors that have been successfully T-dualised giving rise to novel supergravity solutions, the opposite limit, $\lambda \to 0$, would select the topological sector in the T-dual action. Therefore, different $\lambda$ limits connect different types of topological sectors. It would be interesting to better investigate the direct connection between the two. 

In addition, the formal identification between the master action in \eqref{master-action-with-coupling} with $j_\omega = \omega$ and the first order dual action  \eqref{eq:firstorder-index-free}, leads to an alternative use of the $\lambda$ limits. While the $\lambda \to 0 $ regime has still the same interpretation, the opposite regime can be regarded as getting rid of the topological term in \eqref{master-action-with-coupling}, leaving a PCM-like structure, which however lacks a mechanism enforcing the $A$ flatness. It would be interesting to further investigate this issue. 

The above discussion leaves another important question open: what happens in the finite $\lambda$ case, where none of the above two limits is considered? Could one still provide a description of the full T-dual action \eqref{T-dual-sigma model-paper-notation2} in terms of Poisson sigma models and establish a connection with JT theories or, more generally, any other dilaton-gravity theory? 

More generally, it would be interesting to apply a sort of reverse engineering to see if higher dimensional sigma models exist, which under (S)NATD and suitable limits may give rise to non-linear dilaton potentials, or theories with a dynamical dilaton. 

Another interesting question would be to understand which is the analogue of this limiting procedure in the holographically dual field theory.

Finally, we could generalize the SNATD procedure to dS$_3$ or super-dS$_3$ geometries, by performing analytical continuation on the spacetime coordinates. In that case we expect to obtain similar results. 

We hope to come back to some of the above open questions in the near future.

\vskip 30pt

\acknowledgments

We thank Lewis Cole, Pietro Antonio Grassi, Yolanda Lozano, Carlos Nu\~nez, Dima Sorokin, Gabriele Tartaglino Mazzucchelli, Martin Wolf and Salomon Zacarias for useful discussions. SP and AR are partially supported by the INFN grant {\em{Gauge and String Theory (GAST)}}. The work of AR has been supported by INFN-UNIMIB contract, number 125370/2022. The work of DB has been supported by Thailand NSRF via PMU-B, grant number B13F670063. DB thanks the Physics Department at the University of Milano-Bicocca for the warm hospitality during the early stages of this work.

\newpage

\appendix
\section{Conventions}
\label{appendix:osp-algebra}

In this appendix we collect a useful set of conventions that have been used in the main text to describe 3D target supermanifolds. 

\vspace{2mm}
\noindent
\textbf{Spinors.} Locally, the supermanifold is described by a set of three flat bosonic coordinates plus two spinorial ones,
$X^A = (X^a, \theta^\alpha)$, $a=0,1,2, \alpha =1,2$. 

We work in Minkowski signature, with $\eta^{ab}=\eta_{ab}=\text{diag}(-1,+1,+1)$. The Levi-Civita symbol in 3D  is chosen as  $\varepsilon_{012}=-1$ and satisfies the relations
\begin{equation}\label{levi-civita-identities}
\varepsilon_{abc}\varepsilon^{adk}=\delta_{b}{}^{k}\delta_{c}{}^{d}-\delta_{b}{}^{d}\delta_{c}{}^{k} \qquad \qquad \varepsilon_{abc}\varepsilon^{abd}=-2\delta_{c}{}^{d} \,\,.
\end{equation}
Introducing the antisymmetric tensors
\begin{equation}\label{explicit-representation-epsilon}
\epsilon_{\alpha\beta}=
\left( \begin{matrix}
0 & -1 \\
1 & 0
\end{matrix} \right)
\qquad 
\qquad
\epsilon^{\alpha\beta}=
\left( \begin{matrix}
0 & 1 \\
-1 & 0
\end{matrix} \right) \, , 
\end{equation}
which satisfy $\epsilon^{\alpha\gamma}\epsilon_{\gamma\beta}=\epsilon_{\beta\gamma}\epsilon^{\gamma\alpha}=\delta_{\beta}{}^{\alpha}$, indices are raised and lowered as
\begin{equation}
X_a = \eta_{ab} X^b \, , \qquad X^a = \eta^{ab} X_b \, , \qquad \theta_\alpha = \epsilon_{\alpha\beta} \,\theta^\beta  \,  \qquad \theta^\alpha = \epsilon^{\alpha\beta} \theta_\beta \, .
\end{equation}

We make the following choice of gamma matrices 
\begin{equation}\label{explicit-representation-gamma-matrices}
(\gamma^{0})_{\alpha}^{\, \, \, \beta}=
\begin{pmatrix}
0 & 1 \\
-1 & 0
\end{pmatrix} 
\qquad 
(\gamma^{1})_{\alpha}^{\, \, \, \beta}=
\begin{pmatrix}
0 & 1 \\
1 & 0
\end{pmatrix}
\qquad
(\gamma^{2})_{\alpha}^{\, \, \, \beta}=
 \begin{pmatrix}
1 & 0 \\
0 & -1
\end{pmatrix}  \,\, .
\end{equation}
They are subject to the ordinary Clifford algebra and realize the $\mathfrak{sl}(2,\mathbb{R})$ algebra in the fundamental representation, 
\begin{equation}\label{osp(1|2)-def-gamma-matrices}
\{ \gamma^{a},\gamma^{b} \}_{\alpha}{}^{\beta}=2\delta_{\alpha}{}^{\beta}\eta^{ab} \qquad \qquad
[\gamma^{a},\gamma^{b}]_{\alpha}{}^{\beta}=2\varepsilon^{abc}(\gamma_{c})_{\alpha}{}^{\beta} \,\, .
\end{equation}
In addition, the following identity holds
\begin{equation}
 (\gamma^{a})_{\alpha}{}^{\delta}(\gamma^{b})_{\delta}{}^{\beta}=\delta_{\alpha}{}^{\beta}\eta^{ab}+\varepsilon^{abc}(\gamma_{c})_{\alpha}{}^{\beta}  \,\, . 
\end{equation}
If we define $(\gamma^a)_{\alpha\beta} = \epsilon_{\beta \delta} (\gamma^{a})_{\alpha}{}^{\delta}$ and $(\gamma^a)^{\alpha\beta} = \epsilon^{\alpha \delta} (\gamma^{a})_{\delta}{}^{\beta}$, we explicitly obtain
\begin{equation}\label{explicit-representation-upper-index-gamma-matrices}
\begin{aligned}
&(\gamma^{0})^{\alpha\beta}= (\gamma^{0})_{\alpha\beta} = 
\left( \begin{matrix}
-1 & 0 \\
0 & -1
\end{matrix} \right)
\qquad 
(\gamma^{1})^{\alpha\beta}= - (\gamma^{1})_{\alpha\beta} =
\left( \begin{matrix}
1 & 0 \\
0 & -1
\end{matrix} \right) \\
& \qquad \qquad \quad \qquad(\gamma^{2})^{\alpha\beta}=
-(\gamma_{2})_{\alpha\beta} =
\left(  \begin{matrix}
0 & -1 \\
-1 & 0
\end{matrix} \right)
\end{aligned}
\end{equation}
These are symmetric matrices, which enjoy the following property
\begin{equation}\label{gamma-matrix-identity-vector-index-contracted}
(\gamma^{a})_{\alpha\beta}(\gamma_{a})_{\rho\sigma}=\epsilon_{\rho\alpha}\epsilon_{\beta\sigma}+\epsilon_{\rho\beta}\epsilon_{\alpha\sigma} \, .
\end{equation}

\vspace{2mm}
\noindent
\textbf{The $\mathfrak{osp}(1|2)$ superalgebra.} The Lie superalgebra $\mathfrak{osp}(1|2)$ underlies the model studied in this work and includes three bosonic and two fermionic generators, $T_A \equiv (L_a, Q_\alpha)$, $a \in \{0,1,2\}$, $\alpha \in \{ 1,2\}$, obeying commutation relations $[T_A,T_B]=f_{AB}{}^{C}T_C$ given by
\begin{equation}\label{osp-algebra-vector-notation}
[L_{a},L_{b}]=\varepsilon_{ab}{}^{c}L_{c} \qquad 
\{ Q_{\alpha},Q_{\beta} \}=-i(\gamma^{a})_{\alpha\beta}L_{a} \qquad [L_{a},Q_{\alpha}]=-\tfrac{1}{2}(\gamma_{a})_{\alpha}{}^{\beta}Q_{\beta} \, \, .
\end{equation}
The algebra can be recast in double spinor notation by defining
$L_{\alpha\beta}=-i(\gamma^{b})_{\alpha\beta}L_{b}$. In terms of these generators it reads 
\begin{equation}
[L_{\alpha\beta},L^{\gamma\delta}]\ =\ -2{\rm i}{\delta_{(\alpha}}^{(\gamma}{L_{\beta)}}^{\delta)}  \qquad 
\{ Q_{\alpha},Q_{\beta} \}=L_{\alpha\beta} \qquad [L_{\alpha\beta},Q_{\gamma}]=-i\epsilon_{\gamma(\alpha}Q_{\beta)}  \, \, ,
\end{equation}
where brackets denote index symmetrisation. This is the form of the algebra used in \cite{Bielli:2021hzg}.
\indent
The non-degenerate, Ad-invariant and graded symmetric bilinear form is taken to be
\begin{equation}
\label{eq:bilinearform}
\langle T_A, T_B \rangle = \delta_{AB} \, , \qquad \quad 
\delta_{AB} = \begin{pmatrix}
-\frac12 \eta_{ab} & 0 \\
0 & i \epsilon_{\alpha \beta}
\end{pmatrix} \,\, ,
\end{equation}
and its inverse reads
\begin{equation}
\label{eq:delta}
\delta^{AB} = \begin{pmatrix}
-2 \eta^{ab} & 0 \\
0 & i \epsilon^{\alpha \beta}
\end{pmatrix}  \qquad \text{such that} \qquad (-1)^C\delta_{AC}\delta^{CB}=\delta_{A}{}^{B} \,\, .
\end{equation}
Lie algebra valued quantities $U\equiv U^A T_A$ naturally have upper indices $U^A = (U^a, U^\alpha)$ and we define associated lower index quantities by
\begin{equation}
\tilde{U}_A \equiv U^B\delta_{BA} = (-\tfrac{1}{2}U^b\eta_{ba}, iU^{\beta}\epsilon_{\beta\alpha}) = (-\tfrac{1}{2}U_a, -i U_\alpha)
\end{equation}
In the main text we shall remove the tilde from the latter quantities and simply write
\begin{equation}
U^A V_A \equiv U^A \tilde{V}_A  = U^A V^B \delta_{BA} = -\tfrac{1}{2}U^a V_a -i U^\alpha V_\alpha
\end{equation}

\vspace{2mm}
\noindent
\textbf{Worldsheet forms.} We describe the worldsheet $\Sigma$ of the two-dimensional sigma model with a set of local coordinates $\sigma^1 =\tau, \sigma^2 = \sigma$. For elementary differentials $\mathrm{d}\sigma^i$ and their Hodge duals, we define
\begin{equation}
    \mathrm{d}\sigma^i \wedge \mathrm{d}\sigma^j= \mathrm{d} \tau \mathrm{d}\sigma \, \varepsilon^{ji} \, , \qquad \quad 
    \mathrm{d}\sigma^i \wedge \star \mathrm{d}\sigma^j = \mathrm{d} \tau \mathrm{d}\sigma \, \sqrt{-h} \, h^{ij} \, , 
\end{equation}
where $h_{ij}$ is the two-dimensional metric on $\Sigma$, $h^{ij}$ its inverse and $h = \mathrm{det}(h_{ij})$. This easily generalizes to the product of 1-forms $A = \mathrm{d}\sigma^i A_i$ and $B = \mathrm{d}\sigma^j B_j$
\begin{equation}
 A \wedge B =  \mathrm{d} \tau \mathrm{d}\sigma \, \varepsilon^{ij} A_j B_i  
 \, , \qquad \quad 
 A \wedge \star B =  \mathrm{d} \tau \mathrm{d}\sigma \, \sqrt{-h} \, h^{ij} A_j B_i \,\, ,
\end{equation}
while two-forms are defined as
\begin{equation}
C = \mathrm{d}\sigma^{i} \wedge \mathrm{d}\sigma^{j} \, 
 C_{ji} 
\end{equation}

\vspace{2mm}
\noindent
\textbf{Sigma model actions.}
The dual sigma model action can be written in index-free notation, as in \eqref{T-dual-sigma model-paper-notation}, or equivalently in a coordinate formulation as in \eqref{general-T-dual-sigma-model}. In translating from the former to the latter the dual metric $G$ and the $B$ field are respectively identified, up to an overall $-1/2$ factor, as the symmetric and antisymmetric components of the operator $\mathcal{O}\equiv \frac{1}{1-\mathrm{ad}_X}=\sum_{k=0}^\infty \mathrm{ad}_X^k = \mathcal{O}^{\text{(s)}}+\mathcal{O}^{\text{(a)}}$ with respect to the inner product. They read
\begin{equation}
\mathcal{O}^{\text{(s)}} \equiv  \sum_{k=0}^\infty \mathrm{ad}_X^{2k} =\frac{1}{1-\mathrm{ad}_X^2}
\qquad \qquad \qquad
\mathcal{O}^{\text{(a)}} \equiv  \sum_{k=0}^\infty \mathrm{ad}_X^{2k+1} =\frac{\mathrm{ad}_X}{1-\mathrm{ad}_X^2} \,\, ,
\end{equation} 
and exploiting the identities
\begin{equation}
\langle A, \star B \rangle = - \langle \star A, B \rangle = \langle B ,\star A \rangle \qquad \text{and} \qquad \langle A ,\mathrm{ad}_{X}^{k}(B) \rangle = (-1)^{k} \langle \mathrm{ad}_{X}^{k}(A),B \rangle \,\, ,
\end{equation}
for any two Lie algebra valued 1-forms $A,B$, one can indeed observe that
\begin{equation}
\langle A, \mathcal{O}^{\text{(s)}}(B) \rangle = \langle \mathcal{O}^{\text{(s)}}(A), B  \rangle \qquad \qquad \langle A, \mathcal{O}^{\text{(a)}}(B) \rangle = - \langle \mathcal{O}^{\text{(a)}}(A), B  \rangle \,\, .
\end{equation}
These in turn lead to 
\begin{equation}
\langle A, \mathcal{O}^{\text{(s)}}(A) \rangle = 0 \qquad \qquad \qquad \qquad  \langle A, \mathcal{O}^{\text{(a)}}(\star A) \rangle = 0
\end{equation}
and finally to the identifications
\begin{equation}
G = \langle \textrm{d} X, \mathcal{O}^{\text{(s)}}(\star \textrm{d}X) \rangle
\qquad \quad \text{and} \qquad \quad
B =  \langle \textrm{d} X, \mathcal{O}^{\text{(a)}} (\textrm{d}X) \rangle \,\, .
\end{equation}
The relation between the two formulations is then easily determined
\begin{equation}
\begin{aligned}
\langle \textrm{d} X, \mathcal{O}^{\text{(s)}}(\star \textrm{d}X) \rangle & =  \langle \textrm{d} X, \star \textrm{d}X^A \mathcal{O}^{\text{(s)}}{}_{A}{}^{C}T_C \rangle =  \textrm{d} X^B \wedge \star \textrm{d}X^A \mathcal{O}^{\text{(s)}}{}_{A}{}^{C}\delta_{CB}  \\
 & \equiv \mathrm{d}X^B \wedge \star \mathrm{d}X^A G_{AB}(X) = \textrm{d}\tau \textrm{d}\sigma \, \sqrt{-h} \, h^{ij} \, \partial_j X^B \partial_i X^A G_{AB}(X)
\\
\\
\langle \textrm{d} X, \mathcal{O}^{\text{(a)}} (\textrm{d}X) \rangle & = \langle \textrm{d} X, \textrm{d}X^A \mathcal{O}^{\text{(a)}}{}_{A}{}^{C}T_C \rangle =  \textrm{d} X^B \wedge \textrm{d}X^A \mathcal{O}^{\text{(a)}}{}_{A}{}^{C} \delta_{CB}
\\
 & \equiv \mathrm{d}X^B \wedge  \mathrm{d}X^A B_{AB}(X) = \textrm{d}\tau \textrm{d}\sigma \, \varepsilon^{ij} \, \partial_j X^B \partial_i X^A B_{AB}(X) \, .
 \end{aligned}
\end{equation}
Given the close connection between $G,B$ and $\mathcal{O}^{\text{(s)}},\mathcal{O}^{\text{(a)}}$, in the second part of the subsection \ref{subsec:JT_from_NATD} we shall not distinguish between them.

\vskip 10pt

\section{Matching the dual model to sAdS$_2$}
\label{sec:matching}

As mentioned in the main text, we seek to write the dual model \eqref{eq:PCDMPoincaremetric} in the form \eqref{eq:ansatz}. When the transformations \eqref{eq:tildethetaTothetaGeneral} are performed in the sAdS$_2$ metric  \eqref{eq:dsVerlinde}, we obtain
\begin{align}
8&\mathrm{d}s^2_{\text{sAdS}_2} \! =
\label{eq:uplift} \\
& \!-\frac{1}{u^2}\mathrm{d}t^2
\left(1\!-\!2 i \! \left[ c_{12} \left(\partial_t c_{11}\!-\!u^{-1} c_{21}\right)\!+\!c_{11}
\left(u^{-1} c_{22}\!-\!\partial_t c_{12}\right)\!+\! c_{22}\partial_t c_{21}\!-\!c_{21} \partial_t c_{22} \right]\!\theta^1 \theta^2\right)
\notag \\
&\!+\!\frac{\mathrm{d}u^2}{u^2} \left(1\!-\!2 i  u^{-1} \! \left[ c_{11}c_{22}\!-\!c_{12}c_{21}\!-\!u(c_{21}\partial_u c_{12}\!-\! c_{22}\partial_u c_{11}\!+\!c_{11}\partial_u c_{22}\!-\!c_{12}\partial_u c_{21}) \right]\!\theta^1 \theta^2\right)
\notag \\
&\!-\!\frac{2
i}{u^2} \mathrm{d}t \mathrm{d}u \! \left( c_{12}\! \left[  \partial_t c_{21}\!-\!\partial_u c_{11} \right] \!+\!
c_{22}\! \left[ \partial_t c_{11}\!-\!
\partial_u c_{21}  \right] \!-\!c_{11}\!\left[  \partial_t c_{22}\!-\! \partial_uc_{12} \right] \!-\!c_{21} \! \left[  \partial_t c_{12} \!-\! \partial_u c_{22}  \right]  \right)\!\theta^1 \theta^2
\notag \\
&\!-\! \frac{4}{u^2} \mathrm{d}\theta^1 \mathrm{d}\theta^2 \theta^1 \theta^2 (c_{12} c_{21}\!-\!c_{11} c_{22})^2
\notag \\
&\!+\!\frac{2 i}{u^2}
\mathrm{d}\theta^1 \theta^1 \!\left( \mathrm{d}t \left[ c_{11}^2\!+\!c_{21}^2\right] \!-\!2\mathrm{d}u c_{11} c_{21}\right)\!+\!\frac{2
i}{u^2} \mathrm{d}\theta^2 \theta^2 \left( \mathrm{d}t \left[ c_{12}^2\!+\!
c_{22}^2\right] \!-\!2\mathrm{d}u c_{12} c_{22}\right)
\notag \\
&\!+\!\frac{2 i}{u^2}  \left( \left[ c_{11}
c_{12}\!+\!c_{21} c_{22}\right] \mathrm{d}t\!-\! \left[ c_{11} c_{22}\!+\!c_{12} c_{21}\right] \mathrm{d}u\right)(\mathrm{d}\theta^1 \theta^2\!+\!\mathrm{d}\theta^2 \theta^1),
\notag
\end{align}
which is now written in the $\theta$ variables. Substituting the above expression in \eqref{eq:ansatz} leads to a system of eleven functions $\{c_{11},c_{12},c_{21},c_{22},g_1,g_1^\theta,g_2,g_2^\theta,g_3,g_4,g_4^\theta\}$ to be determined.  

First of all, matching the two expressions \eqref{eq:ansatz} and \eqref{eq:PCDMPoincaremetric} we fix the $g_i$ and $g_i^\theta$ functions, which at this level are still functions of $c_{ij}$ and their derivatives.  
Substituting the $g_i$, $g_i^\theta$ functions and matching the rest of the components, we obtain two differential equations
\begin{equation}
\begin{aligned}\label{eq:ODE}   
&(c_{12}\partial_uc_{21}\!-\!c_{11}\partial_uc_{22})\!-\!c_{12}\partial_tc_{11}\!+\!c_{11}\partial_tc_{12}\!-\!c_{22}(\partial_tc_{21}-\partial_uc_{11}\!)\!+\!c_{21}(\partial_tc_{22}-\partial_uc_{12}\!)\!=\!0,
\\
&(c_{22}\partial_uc_{21}\!-\!c_{21}\partial_uc_{22})\!-\!c_{22}\partial_tc_{11}\!+\!c_{21}\partial_tc_{12}\!-\!c_{12}(\partial_tc_{21}-\partial_uc_{11}\!)\!+\!c_{11}(\partial_tc_{22}-\partial_uc_{12}\!)\!=\!0,
\end{aligned}
\end{equation}
and six algebraic ones
\begin{equation}
    \begin{split}
        \label{eq:algebraicequations}
    &c_{11}c_{21}-tu=0\\
    &c_{12}c_{22}=0\\
    &c_{12}c_{21}+c_{11}c_{22}-u=0\\
    &t-(c_{11}c_{12}+c_{21}c_{22})=0\\
    &u^{2}+t^2-(c_{11}^2+c_{21}^2)=0\\
    &c_{12}^2+c_{22}^2-1=0 \, . 
    \end{split}
\end{equation}
The first four equations in \eqref{eq:algebraicequations} can be easily solved, leading to two sets of solutions,
\begin{align}
\label{eq:hsSols}
&\text{Solution } 1: \; \;\{c_{11},c_{12},c_{21},c_{22}\}=\pm\{t,\;1\;,u\;,0\},\\
 \label{eq:hsSols2}
 &\text{Solution } 2: \; \;\{c_{11},c_{12},c_{21},c_{22}\}=\pm\{u,\;0,\;t,\;1\}.
\end{align}
These in turn satisfy automatically the last two algebraic equations in \eqref{eq:algebraicequations} and the differential equations \eqref{eq:ODE}.
Exploiting Solution $1$ in \eqref{eq:tildethetaTothetaGeneral} one obtains the explicit transformations given in \eqref{eq:tildethetaTothetaExplicit} and the dual supermetric \eqref{eq:sol2}  written in terms of the sAdS$_2$ metric. The discussion of Solution $2$ is given in the next subsection.

\vspace{2mm}
\noindent
\textbf{The second solution.} Referring to the general transformation \eqref{eq:tildethetaTothetaGeneral}, Solution $2$ \eqref{eq:hsSols2} corresponds to the following  change of fermionic  coordinates 
\begin{align}
\label{eq:tildethetaTothetaExplicit2}
\tilde{\theta}^1=\;u\theta^1,\quad\qquad \tilde{\theta}^2=\;t\theta^1+\theta^2\, ,
\end{align}
while the associated $G_i$ functions in \eqref{eq:ansatz} read
\begin{equation}
\begin{alignedat}{3}  &G_1(r,\theta)&&=\frac{2r^2}{r^2-1}+ i \frac{4r^2(4r^2-7)}{3(r^2-1)^2} \theta^1\theta^2 \, , &&\qquad \quad
G_3(r,\theta)=-\frac{ir(r^2-8)}{6(r^2-4)} + g_3^\theta (r) \theta^1\theta^2 \, , \\ &G_2(r,\theta)&&=\frac{1}{4}-\frac{i(r^4-4r^2-16)}{6(r^2-4)^2} \theta^1\theta^2 \, , &&\qquad \quad
    G_4(r,\theta)= \frac{2i}{3}+\frac{r^2+12}{18} \theta^1\theta^2 \, ,
\end{alignedat}
\end{equation} 
 Using these entries, the dual metric and the $B$ field become
\begin{align}
\label{eq:sol1}
    &\mathrm{d}s^2=\frac{2r^2}{r^2-1}\!\left(1+\frac{2i(4r^2-7)}{3(r^2-1)}\theta^1\theta^2 \! \right) \! \mathrm{d}s^2_{\text{sAdS}_2} 
    \\
    &+\! \frac{1}{4}\!\left(\!1\!-\!\frac{2i(r^4-4r^2-16)}{3(r^2\!-\!4)^2}\theta^1\theta^2\!\right)\!\mathrm{d}r^2\!-\!\frac{ir(r^2-8)}{6(r^2\!-\!4)}\mathrm{d}(\theta^1\theta^2)\mathrm{d}r\!+\frac{2i}{3}\!\left( \! 1\! -\! \frac{i(r^2+12)}{12} \!  \theta^1\theta^2\! \right) \!\mathrm{d}\theta^1\mathrm{d}\theta^2\! ,
    \nonumber \\
    \nonumber \\
   &B=-\frac{2r^3}{r^2\!-\!1}\!\left(\!1\!+\!\frac{i(11r^2\!-\!17)}{3(r^2\!-\!1)}\theta^1\theta^2\!\right)\!B_{\text{sAdS}_2}
   \label{eq:sol1B2} \\
   &\quad\!-\!\frac{4i}{3(r^2\!-\!4)}\!\mathrm{d}r\wedge\mathcal{A}\!+\!\frac{ir}{3}\!\left(\!1-\!\frac{10i}{3}\theta^1\theta^2\!\right)\!\left(\!\mathrm{d}_B\mathcal{A}\!-\!\frac{1}{2}\mathrm{d}_F\mathcal{A}\!\right),
   \nonumber
\end{align}
where the $\mathrm{d}s^2_{\text{sAdS}_2}$ and $B_{\text{sAdS}_2}$ are  
\begin{align}
\label{eq:dsVerlindesol1}
\mathrm{d}s^2_{\text{sAdS}_2}=&\frac{1}{8}\!\left(1\!-\!4i\theta^1\theta^2\right)\mathrm{d}s^2_{\text{AdS}_2}\!-\!\frac{1}{2}\theta^1 \theta^2\mathrm{d}\theta^1 \mathrm{d}\theta^2\!-\!\frac{i}{4}\!\left(\mathrm{d}u\partial_t\!+\!\mathrm{d}t\partial_u\right)\!\mathcal{A},\\
\label{eq:B2Verlindesol1}
B_{\text{sAdS}_2}=&-\frac{1}{4}\!\left(1\!-\!4i\theta^1\theta^2\right)\!\left({\rm vol}_{\text{AdS}_2}+\theta^1\theta^2 \mathrm{d}_F\mathcal{A}-i\mathrm{d}_B\mathcal{A}\right).
\end{align}

\vspace{2mm}
\noindent
\textbf{Dualisation of the maximal bosonic subgroup.} When dualising the SL$(2,\mathbb{R})_{L}$ subgroup of 
$\text{OSp}(1|2)_{L} \times \text{OSp}(1|2)_{R}$, the matching between the T-dual model and the transformed Verlinde metric requires solving exactly the same differential and algebraic equations, as \eqref{eq:ODE} and \eqref{eq:algebraicequations}. Therefore, once again we find the same two sets of solutions
\begin{align}
\label{eq:hsSolsBoso}
&\text{Solution } 1: \; \;\{c_{11},c_{12},c_{21},c_{22}\}=\pm\{t,\;1\;,u\;,0\},\\
 \label{eq:hsSolsBoso2}
 &\text{Solution } 2: \; \;\{c_{11},c_{12},c_{21},c_{22}\}=\pm\{u,\;0,\;t,\;1\}.
\end{align}
In the main text we discussed Solution 1. Here, for completeness, we present Solution 2. The corresponding functions defined in \eqref{eq:ansatz} are
\begin{equation}
\begin{aligned}
&G_{1}(r,\theta)=\frac{2r^2}{r^2-1}(1+4i\theta^1\theta^2), \qquad \qquad \quad G_{3}(r,\theta)=g_3^\theta (r)\theta^{1}\theta^{2}, 
\\
& G_{2}(r,\theta)=\frac{1}{4}, \qquad\qquad\qquad\qquad\qquad\qquad \,\, G_{4}(r,\theta)=2i+2\theta^{1}\theta^{2}.
\end{aligned}
\end{equation}
The T-dual metric and B-field read
\begin{equation}
\mathrm{d}s^2 = \frac{2r^2}{r^2-1}(1+4i\theta^1\theta^2)\mathrm{d}s^2_{\text{sAdS}_2}+\frac{1}{4}\mathrm{d}r^2+2i(1-i\theta^1\theta^2)\mathrm{d}\theta^1 \mathrm{d}\theta^2,
\end{equation}
\begin{equation}
B=-\frac{2r^3}{r^2-1}(1+4i\theta^1\theta^2)B_{\text{sAdS}_2}+\mathrm{d}\left[\frac{ir\mathcal{A}}{2}\right]-\frac{ir}{2}(1-i\theta^1\theta^2)\mathrm{d}_{F}\mathcal{A}
%B=-\frac{2r^3}{r^2-1}(1+4i\theta^1\theta^2)B_{\text{sAdS}_2,\theta} -\frac{r(2r^2+1)}{4(r^2-1)}\theta^1 \theta^2 \mathrm{d}_{F}\mathcal{A}+\frac{ir}{2}\mathrm{d}_{B}\mathcal{A} + \frac{i}{2} \mathrm{d}r \wedge \mathrm{d}\mathcal{A} \, , 
\end{equation}
with $\mathrm{d}s^2_{\text{sAdS}_2}$ and $B_{\text{sAdS}_2}$ given in \eqref{eq:dsVerlindesol1} and \eqref{eq:B2Verlindesol1}.

\newpage
\bibliography{Bibliography}

\end{document}